# Chemical Programming of Stacking Order and Superconductivity Dimensionality in $NbSe_{2-x}Te_x$ Polytypic Heterostructures

Xiangqi Liu[1†], Haonan Wang[2†], Ziguang Liu[1†], Peihan Sun[3†], Xiang Xu[1,8], Jing Jiang[4,5], Zhengyang Li[1], Ziyi Zhu[6], Wei Xia[1,8], Xia Wang[1,7], Leiming Chen[6], Ke Qu[2], Yulin Chen[1,8], Cheng Chen[1,8*], Kai Liu[4,5*], Zhenzhong Yang[2*], Yanfeng Guo[1,8*]

[1]*State Key Laboratory of Quantum Functional Materials, School of Physical Science and Technology, ShanghaiTech University, Shanghai 201210, China*

[2]*Key Laboratory of Polar Materials and Devices (MOE), Department of Electronics, School of Information and Electronic Engineering, East China Normal University, Shanghai 200241, China*

[3]*Department of Physics, School of Science, Hebei University of Science and Technology, Shijiazhuang 050018, China*

[4]*School of Physics and Beijing Key Laboratory of Optoelectronic Functional Materials & MicroNano Devices, Renmin University of China, Beijing 100872, China*

[5]*Key Laboratory of Quantum State Construction and Manipulation (Ministry of Education), Renmin University of China, Beijing 100872, China*

[6]*Henan Key Laboratory of Aeronautical Materials and Application Technology, Zhengzhou University of Aeronautics, Zhengzhou, Henan 450046, China*

[7]*Analytical Instrumentation Center, School of Physical Science and Technology, ShanghaiTech University, Shanghai 201210, China*

[8]*ShanghaiTech Laboratory for Topological Physics, ShanghaiTech University, Shanghai 201210, China*

[†]These authors contributed equally to this work.

[*]Corresponding authors: chencheng1@shanghaitech.edu.cn; zzyang@phy.ecnu.edu.cn; kliu@ruc.edu.cn; guoyf@shanghaitech.edu.cn.

## ABSTRACT

**Layered transition-metal dichalcogenides exhibit electronic states highly sensitive to lattice symmetry, yet elucidating the structure–superconductivity relationship has been challenging due to the lack of a continuous structural evolution within a single material system. Here, we establish $NbSe_{2-x}Te_x$ as a model system, where Te substitution drives a sequential polytypic transition across the full composition range—from the low-Te 2H phase, through mixed 1H/1T heterostructured 4H and 6R polytypes at intermediate substitution, to ideal 1T and distorted 1T' structures at high Te content. Combining comprehensive characterizations, we map both the structural evolution and the accompanying changes in spectral weight and orbital character near the Fermi level. Electrical transport measurements reveal a striking phase-dependent superconducting dimensionality: the 4H phase ($NbSe_{1.4}Te_{0.6}$) exhibits enhanced anisotropy and an anomalous boost in the in-plane upper critical field, while the 6R phase ($NbSe_{1.2}Te_{0.8}$) shows suppressed $T_c$ (~ 2.8 K) and three-dimensional-like behavior. The combined electronic-structure and superconducting results connect the stacking sequence to effective interlayer coherence and the local spin-orbit environment. SOC band**

**calculations resolve the distinct symmetry constraints on the 4H and 6R electronic states. These mechanisms fundamentally differ from conventional approaches by leveraging intrinsic, composition-driven polytypism to modify interlayer coupling and band structure. Thus, by establishing a definitive phase diagram that directly correlates polytypism with superconducting dimensionality, our work provides a clear roadmap for the rational design of low-dimensional superconducting devices through stacking-sequence engineering.**

## INTRODUCTION

Layered transition-metal dichalcogenides (TMDs) represent a pivotal platform for exploring the intricate coupling between crystal structure and collective electronic states. Their covalently bonded X-M-X monolayers, where M is a transition metal and X is a chalcogen, are stacked along the *c*-axis via weak van der Waals (vdW) interactions. This unique anisotropic architecture allows external perturbations, such as chemical doping, to readily tune the electronic structure or drive structural phase transitions, thereby dramatically reshaping correlated electronic ordered states including superconductivity and charge-density-wave (CDW) order[1-4].

Among superconducting TMDs, $NbSe_2$ stands out as a prototypical system. In its bulk form, $NbSe_2$ stabilizes in the 2H phase featuring a trigonal-prismatic coordination of Nb atoms, exhibiting coexisting CDW order and superconductivity with a transition temperature $T_c$ of approximately 7.2 K[5-8]. Although thinning 2H-$NbSe_2$ down to the monolayer limit breaks inversion symmetry and induces robust Ising superconducting protection against Pauli depairing, the structural and electronic tunability within bulk crystals remains inherently constrained[9-14]. Replacing Se with the larger Te atom provides a powerful isovalent strategy to overcome these limitations. Due to the differences in atomic radius and electronegativity, Te substitution is expected to systematically modify bond lengths, local coordination geometries, and the interlayer vdW gap through substitution on the chalcogen sites. These structural modifications successfully alter the carrier concentration and Fermi-surface topology, ultimately narrowing the formation-energy gap between distinct coordination geometries and driving structural transitions from the 2H phase to 1T-type phases or self-assembled 1T/1H superlattices[15-21].

Previous studies have highlighted the profound impact of such substitution-induced polytypism. For instance, Te substitution in $TaSe_{2-x}Te_x$ stabilizes multiple polymorphs, where the 3R polytype achieves a $T_c$ up to 2.4 K[15], which is nearly 17 times higher than that of pristine 2H-$TaSe_2$. More recently, investigations into the $NbSe_{2-x}Te_x$ system demonstrated that adjusting the Se/Te stoichiometric ratio reduces the formation energy difference between the 1T and 1H phases, enabling the self-assembly of alternating 1H and 1T layers into bulk heterostructured superlattices that retain the electronic characteristics of both constituent sublattices[16,17]. However, several critical questions remain open. The exact compositional windows that stabilize the different phases lack systematic clarification within a unified study. Superconducting 1T-NbSeTe single crystals and self-assembled 6R H/T superlattices have already been reported[16,17,22]. The comparison of structurally resolved 4H and 6R crystals within one substitution series offers a further opportunity to examine how distinct mixed-layer sequences affect the superconducting field response. Consequently, it remains unresolved whether these heterostructured polytypes merely reflect an average crystal structure or if they directly dictate superconducting dimensionality by modifying interlayer Josephson coupling, disorder scattering, and spin-orbit-induced pair protection.

In this study, we address these outstanding gaps by systematically investigating the structural phase transitions and corresponding superconducting properties of $NbSe_{2-x}Te_x$ single crystals across the entire substitution range of $0 \leq x \leq 2$. High-quality single crystals were successfully synthesized via a two-step chemical vapor transport (CVT) method. Utilizing a comprehensive combination of X-ray diffraction (XRD), atomic-resolution scanning transmission electron microscopy (STEM), and angle-resolved photoemission spectroscopy (ARPES), we map the nominal composition regimes of the polytypes and track the underlying evolution of the electronic band structures, Fermi surfaces, and spectral weights near the Fermi level ($E_F$). Furthermore, transport, magnetization, and specific-heat measurements reveal a striking, phase-dependent evolution of superconducting dimensionality. Our results show that the 4H polytype is associated with enhanced two-dimensional (2D) superconducting behavior and an anomalous in-plane upper critical field $B_{c2}^{//}$ enhancement, consistent with reduced interlayer coherence, whereas the translated stacking sequence of the 6R polytype supports a more 3D superconducting response relative to 4H. These findings establish self-assembled 1H/1T TMD superlattices as a highly tunable bulk platform for engineering low-dimensional quantum devices.

## EXPERIMENTAL SECTION

**Synthesis of $NbSe_{2-x}Te_x$ single crystals.** High-quality $NbSe_{2-x}Te_x$ single crystals were synthesized using a two-step CVT method. Initially, $NbSe_{2-x}Te_x$ polycrystalline powder was prepared as a precursor. High-purity Nb powders (Macklin, 99.99%), Se powders (Macklin, 99.99%), Te powders (Macklin, 99.99%) were mixed in a molar ratio of 0.9: 2-*x*: *x* and placed into an evacuated quartz tube, which was then sealed. This molar ratio specifies the precursor mixture for CVT growth. The mixture was heated to 973 K over a period of 10 hours and maintained at this temperature for 50 hours. Subsequently, the $NbSe_{2-x}Te_x$ polycrystalline powder was combined with iodide (Macklin, 99.99%) in a weight ratio of 1: 0.1, and sealed into 14 cm quartz ampoules after evacuating to a pressure of less than $10^{-4}$ Torr. The reaction zone was then programmed at a higher temperature of 850°C with the growth zone at a lower temperature of 800°C for 15 days, and then quenched in ice-water to avoid mixed phases. Single crystals of $NbSe_{2-x}Te_x$ polytypes were obtained from the crucible.

**Crystal characterizations.** The composition of $NbSe_{2-x}Te_x$ single crystals was analysed using scanning electron microscopy (SEM) equipped with energy-dispersive X-ray spectroscopy (EDS) performed on Phenom Pro (Phenom-World BV). Room-temperature powder X-ray diffraction (PXRD) patterns were collected from $NbSe_{2-x}Te_x$ powder derived from being crushed of single-crystalline samples using a Bruker D8 Venture diffractometer equipped with a Cu K$\alpha$ radiation source ($\lambda$ = 1.5418 Å). Additionally, single-crystal X-ray diffraction (SXRD) measurements were conducted using a Bruker D8 single-crystal X-ray diffractometer with a Mo K$\alpha$ radiation source ($\lambda$ = 0.71073 Å) at room temperature.

**Magnetization and electrical transport measurements.** Magnetic characterization was carried out using a Quantum Design MPMS-3 magnetometer. Temperature-dependent magnetization $M(T)$ was measured under a small magnetic field (varying 5-20 Oe) applied perpendicular to the *ab*-plane, in both zero-field-cooling (ZFC) and field-cooling (FC) modes. Isothermal magnetization curves were acquired across a field range (± 600 Oe) at multiple temperatures. In-plane electrical transport measurements were performed using a standard four-wire method with an excitation current of 100 μA in the PPMS AC drive mode, using a DynaCool Physical Properties Measurement System (PPMS-14T, Quantum Design) equipped with a dilution refrigerator. The AC drive reverses the current polarity and combines the voltage readings to suppress DC offsets. The specific heat of the samples was measured using a standard thermal relaxation technique in a commercial PPMS-14T system equipped with a dilution refrigerator. The background signal was carefully subtracted within the temperature range from 300 K down to 0.06 K to ensure accuracy and reliability of the specific heat data. For the superconducting critical-field measurements, the current was applied within the *ab* plane. The normal-state *c*-axis resistivity $\rho_{zz}$ was measured in a Corbino geometry with current along $c$. Its ratio to the in-plane resistivity $\rho_{xx}$ was compared for representative 2H, 4H and 6R crystals (Fig. S23)[23]. The field orientations $B//ab$ and $B//c$ denote magnetic fields parallel and perpendicular to the layers, respectively. Angular measurements used $\theta = 0°$ for $B//c$ and $\theta = 90°$ for $B//ab$ and were performed at 1.8 K; the additional 4H enhancement is confined to approximately $|\theta - 90°| < 2°$. Fixed-orientation measurements on the representative 6R crystal used the dilution-refrigerator configuration. The sweeps displayed in Fig. 2(i) span 1.8–2.3 K in 0.05 K steps. For transport analysis, the normal-state reference was $R_N = R_{xx}(10\ K)$. The zero-field transition temperature $T_c$ was defined by $R_{xx}(T_c, B = 0) = 0.5R_N$, and critical fields were extracted using the same 50% $R_N$ criterion. The $T/T_c$ plots use this zero-field transition temperature for the corresponding specimen. For anisotropic GL analysis, $B_{c2}^{\perp} = \Phi_0/(2\pi\xi_{ab}^2)$ and $B_{c2}^{//} = \Phi_0/(2\pi\xi_{ab}\xi_c)$; thus the GL fit lengths for $B//ab$ and $B//c$ are $\sqrt{\xi_{ab}\xi_c}$ and $\xi_{ab}$, respectively.

**Cs-corrected STEM measurements.** Cross-sectional transmission electron microscopy (TEM) specimens were prepared using a focused ion beam–scanning electron microscope (FIB-SEM; Thermo Fisher Scientific Helios G4 UX). Atomic-resolution high-angle annular dark-field (HAADF) STEM imaging was performed using a Cs-corrected microscope (JEOL JEM-ARM300F) operated at 300 kV.

**ARPES measurements**

ARPES measurements were performed at Beamline 03U of the Shanghai Synchrotron Radiation Facility (SSRF)[24], using a Scienta Omicron DA30 electron analyzer. The samples were all cleaved and measured at a base pressure better than $4\times10^{-10}$ Torr and a temperature of 15 K. To map the overall band structure, spectra were collected with a photon energy of 80 eV under LH polarization. The overall energy resolution was ~ 30 meV, and the angular resolution was 0.1°.

**Density functional calculations**

All first-principles calculations were performed by using the projector augmented-wave method[25,26] as implemented in the Vienna ab initio simulation package (VASP)[27 - 29]. The generalized gradient approximation (GGA) in the scheme of Perdew-Burke-Ernzerhof[30] was adopted for the exchange-correlation functional. The kinetic energy cutoff of the plane-wave basis was set to 500 eV. The Gaussian smearing method with a width of 0.05 eV was used for the Fermi surface broadening. For all structures, both lattice constants and internal atomic positions were fully relaxed until the forces on all atoms were smaller than 0.01 eV/Å. The DFT-D3 method[31] was used to account for the van der Waals interaction. Structural relaxation and the configurational energy comparison were performed without spin-orbit coupling (SOC). SOC was included in subsequent electronic-structure calculations for 4H-$NbSe_{1.4}Te_{0.6}$ and 6R-$NbSe_{1.3}Te_{0.7}$, using the virtual crystal approximation for Se/Te mixing. The band and spin-projection comparisons were evaluated in the $k_z = 0$ and $k_z = \pi/c$ planes (Figs. S19-S22). The 22 × 22 × 10, 22 × 22 × 5, 22 × 22 × 4, 22 × 22 × 20, and 5 × 6 × 8 Monkhorst-Pack $k$ meshes were used for the Brillouin zone sampling of the primitive cell for the 2H, 4H, 6R, 1T and distorted 1T' phases, respectively. For each type of $NbSe_{2-x}Te_x$, the Te concentration was chosen to be $x = 0$, 0.33, 0.5, 0.83, 1.0, 1.5 and 2.0. All the configurations were generated through the enumeration function in the alloy-theoretic automated toolkit (ATAT) software[32], which can enumerate all inequivalent structures where Se atoms are substituted for Te atoms based on the crystal symmetry under the given supercell size and doping concentration. The finite-supercells sizes for 2H, 4H and 6R phases are all $\sqrt{3} \times \sqrt{3} \times 1$, while the 1T phase is $2\sqrt{3} \times 2\sqrt{3} \times 1$. As for the 1T' phase, considering that its primitive cell already contains 54 atoms and the symmetry of the lattice is relatively low, we carried out the element substitution based on this structure without expanding the unit cell.

## Upper critical field and 2D superconductivity

In this section, we delineate the formulae used to fit the angular and temperature dependences of the upper critical magnetic fields for fields perpendicular and parallel to the sample plane.

To fit the temperature dependence results, we used the Ginzburg Landau description of 3D and 2D superconductivity[33,34]:

For 3D superconductivity:

$$B_{c2} = \frac{\Phi_0}{2\pi\xi_0^2}(1 - T/T_c) \ . \tag{1}$$

For 2D superconductivity:

$$B_{c2} = \frac{\sqrt{3}\Phi_0}{\pi\xi_0 d_{sc}}(1 - T/T_c)^{1/2} \ , \tag{2}$$

where $\Phi_0$ represents the flux quantum, $\xi_0$ is the zero-temperature GL coherence length, $d_{sc}$ denotes the 2D superconducting thickness.

In 4H-$NbSe_{2-x}Te_x$, substitutional Se/Te disorder and stacking faults provide scattering channels that affect the superconducting field response. The Werthamer-Helfand-Hohenberg (WHH) theory describes the upper critical field of a dirty-limit type-II superconductor by incorporating orbital depairing, Pauli paramagnetism and spin-orbit scattering[35-38]. Spin-orbit scattering reduces the effectiveness of paramagnetic depairing and can enhance the parallel critical field. Its dimensionless scattering parameter is distinct from the intrinsic SOC energy scale of the electronic bands. The temperature dependence of $B_{c2}^{//}(T)$ is described by:

$$\ln\left(\frac{1}{t}\right) = \left(\frac{1}{2} + \frac{i\lambda_{so}}{4\gamma}\right)\psi\left(\frac{1}{2} + \frac{\bar{h} + \frac{\lambda_{so}}{2} + i\gamma}{2t}\right) + \left(\frac{1}{2} - \frac{i\lambda_{so}}{4\gamma}\right)\psi\left(\frac{1}{2} + \frac{\bar{h} + \frac{\lambda_{so}}{2} - i\gamma}{2t}\right) - \psi(1/2) \ , \tag{3}$$

where $t = T/T_c$, $\bar{h} = 4B_{c2}/[\pi^2 T_c \ |(dB_{c2}/dT)$ at $T = T_c|]$, and $\gamma = \sqrt{(\alpha\bar{h})^2 - (\lambda_{so}/2)^2}$. The Maki parameter $\alpha = \sqrt{2}B_{orb}(0)/B_P$ characterizes paramagnetic relative to orbital limiting, and $\lambda_{so} = 2\hbar/(3\pi k_B T_c \tau_{so})$ characterizes spin-orbit scattering in the WHH convention. Here $\tau_{so}$ is the spin-orbit scattering time; $\lambda_{so}$ is distinct from an intrinsic band-splitting energy. For the representative 4H-$NbSe_{1.4}Te_{0.6}$ crystal, the fit gives $\alpha$ = 5.2 and $\lambda_{so}$ = 7.1.

The dimensional characteristics of polytypes $NbSe_{2-x}Te_x$ are further elucidated by the angular dependence of the upper critical field, $B_{c2}(\theta)$, as shown in Fig. 2. Initially, we model this behavior using the anisotropic GL model:

$$\left(\frac{B_{c2}(\theta)\sin\theta}{B_{c2}^{//}}\right)^2 + \left(\frac{B_{c2}(\theta)\cos\theta}{B_{c2}^{\perp}}\right)^2 = 1, \quad (4)$$

where $B_{c2}^{//}$ and $B_{c2}^{\perp}$ represent the critical fields for magnetic fields oriented parallel ($\theta$ = 90°) and perpendicular ($\theta$ = 0°) to the layers, respectively. However, given the system's pronounced 2D nature, we also apply the 2D Tinkham model:

$$\left(\frac{B_{c2}(\theta)\sin\theta}{B_{c2}^{//}}\right)^2 + \left|\frac{B_{c2}(\theta)\cos\theta}{B_{c2}^{\perp}}\right| = 1. \quad (5)$$

The two models diverge significantly as the field approaches the layer plane ($\theta \rightarrow$ 90°), with the Tinkham model predicting a sharper enhancement of $B_{c2}$ within the plane, which is a feature corroborated by our experimental data.

## Superconducting gap symmetry analysis

To elucidate the superconducting gap structure, we employ the BCS framework to model the electronic specific heat[39,40]:

$$\gamma_e - \gamma_n = \frac{4N(E_F)}{k_B T^3}\int_0^{+\infty}\int_0^{2\pi}\frac{e^{\zeta/k_B T}}{(1+e^{\zeta/k_B T})^2}\times\left[\varepsilon^2 + \Delta^2(T,\theta) - \frac{T}{2}\frac{d\Delta^2(\theta,T)}{dT}\right]d\theta d\varepsilon - \gamma_n, \quad (6)$$

where $\gamma_e = C_e/T$ and $\zeta = \sqrt{\varepsilon^2 + |\Delta(T,\theta)|^2}$. The *s*- and *d*-wave electronic-specific-heat responses are calculated separately from the BCS integral in Eq. (6), with gap functions $\Delta_s(T)$ and $\Delta_d(T)\cos(2\theta)$, respectively. The two-component *s*+*d* model combines these responses as $C_e(T) = wC_{e,s}(T) + (1 - w)C_{e,d}(T)$. Here $C_{e,s}$ and $C_{e,d}$ are component curves normalized to the same normal-state Sommerfeld coefficient $\gamma_n$, and $0 \leq w \leq 1$ is the fractional electronic weight of the *s*-wave component; $1 - w$ is the *d*-wave weight. Figure S17(d) displays $C_e/T - \gamma_n$, with the normal-state coefficient subtracted from the combined response. This additive construction follows the multicomponent specific-heat framework[40,41]. The anomaly near 2.1 K sets $T_c$ for the thermodynamic analysis. Each component gap amplitude follows the BCS temperature dependence in Eq. (7):

$$\Delta(T) = \Delta_0 \tanh\left[1.74\sqrt{(T_c/T - 1)}\right]. \tag{7}$$

## RESULTS AND DISCUSSION

Fig. 1(a) systematically illustrates the structural evolution of the $NbSe_{2-x}Te_x$ single crystals across typical tellurium concentrations. EDS identifies Nb, Se and Te in the grown crystals and tracks the changing Se/Te composition, with a compositional uncertainty within 10% (Fig. S1). The representative compositions are consistent with the phase assignments from diffraction and STEM. At low Te content, the crystals retain the 2H structure, reflecting the high phase stability of 2H phase at low substitution levels. As the tellurium content further increases, the diminishing energy gap between the H-phase and T-phase configurations promotes the regular assembly of uniformly distributed 1H/1T heterostructures[15-18]. Specifically, within the composition range of $x \approx 0.4$–0.6, the system stabilizes in a hexagonal 4H polytype, refined under the space group $P6_3/mmc$ (No. 194). The unit cell of this 4H phase is built from 1H/1T heterostructural blocks arranged in an antiparallel sequence, forming a four-layer repeat superlattice. The antiparallel sequence is compatible with the centrosymmetric $P6_3/mmc$ structure, in which neighboring blocks are related by the crystallographic stacking operations. For tellurium concentrations spanning $x \approx 0.6$–0.9, the $NbSe_{2-x}Te_x$ crystals transition into a rhombohedral 6R polytype with the space group

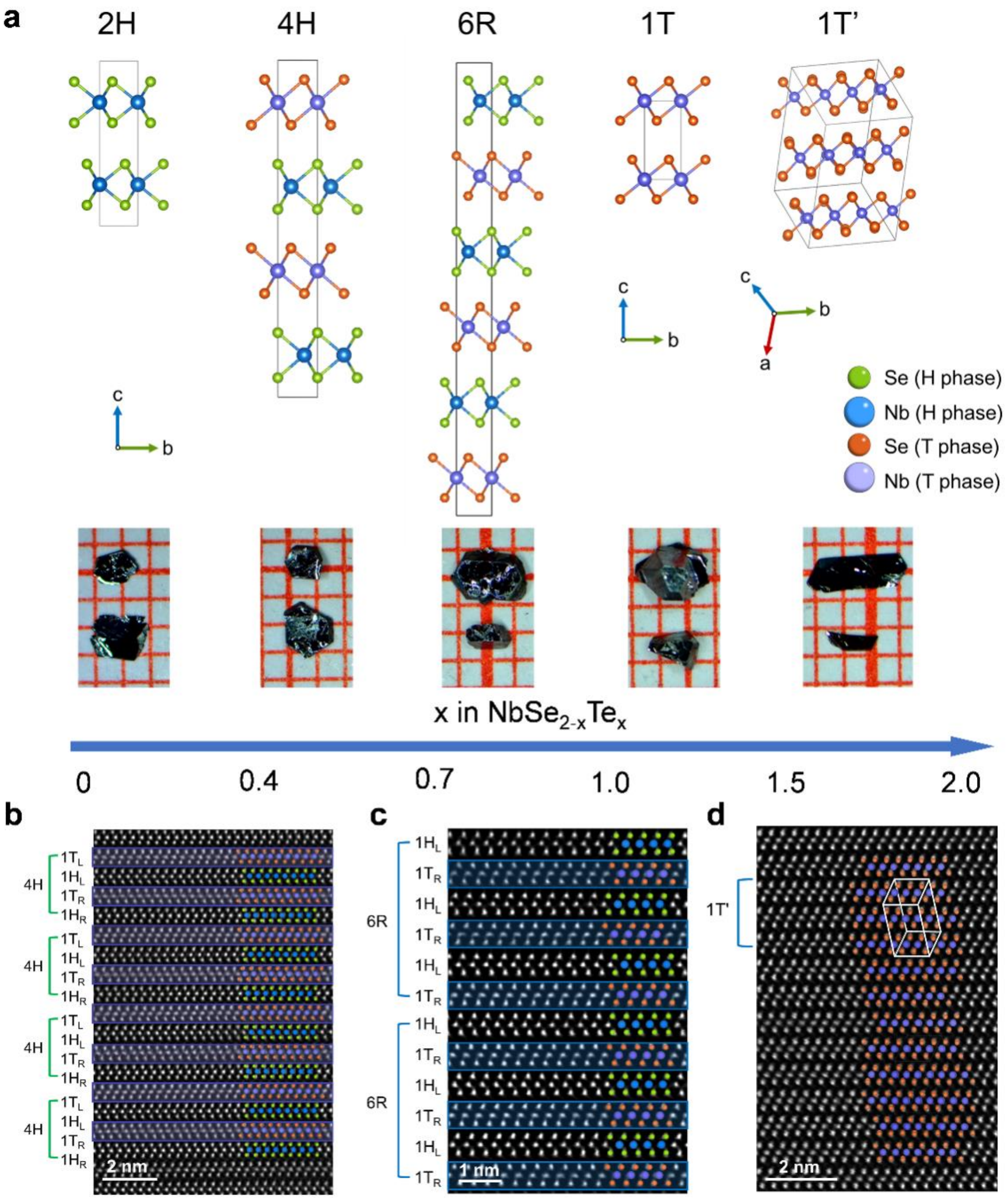

**Figure 1.** (a) Crystal structures and optical images of 2H-$NbSe_2$, 4H-$NbSe_{1.6}Te_{0.4}$, 6R-$NbSe_{1.2}Te_{0.8}$, 1T-NbSeTe, and monoclinic $NbSe_{0.5}Te_{1.5}$, respectively. In H-phase layers, Nb and Se (or site sharing Te) atoms are represented as blue and green balls, respectively; in T-phase layers, Nb and Se atoms are depicted as purple and yellow balls, respectively. (b, c) Cross-sectional STEM images of the 4H and 6R 1H/1T superlattices, respectively, along the [110] direction. To distinguish layer stacking along the *c*-axis, 1H and 1T layers were classified by their in-plane symmetry. Specifically, the tilt direction (left or right) of trigonal prisms (1H) and octahedra (1T) denoted as $1H_L/1H_R$ and $1T_L/1T_R$, respectively. The alternating arrangement of the 1H/1T heterostructures are superimposed with the corresponding atomic structure model. (d) Cross-sectional STEM image of a distorted 1T'-$NbSe_{0.5}Te_{1.5}$ single crystal, overlaid with the corresponding atomic structural model. The crystallographic unit cell is outlined in white.

*R*3*m* (No. 160). In this 6R phase, the 1H/1T bilayers are translated in an A-B-C-A sequence along the *c*-axis. This translated *R*3*m* sequence has a different global symmetry from the centrosymmetric 4H stack. At higher tellurium concentrations ($x > 0.9$), these heterostructured polytypes give way to pure T-coordinated configurations, initially stabilizing in an ideal 1T structure with space group *P*-3*m*1 (No. 164) over the range of $0.9 < x < 1.3$. Beyond $x \sim 1.3$, a further phase transition leads to the dominance of a monoclinic distorted 1T' phase belonging to the space group *C*2/*m* (No. 12). The distorted octahedral coordination within this monoclinic lattice expands the crystallographic unit cell while maintaining an overall centrosymmetric structure. This high-Te distorted regime extends to $NbTe_2$ ($x = 2$), whose monoclinic *C*2/*m* distorted-1T structure is established by diffraction and is denoted 1T' here[42].

To resolve the atomic-scale coordination and layer-stacking sequences, cross-sectional atomic-resolution STEM imaging was performed along the [110] zone axis (Fig. 1(b) and Figs. S3-S4). Fig. 1(b) shows a representative atomic resolution STEM image of the 4H-$NbSe_{1.6}Te_{0.4}$ polytype, in which the alternating 1H and 1T layers are clearly distinguished. To systematically define the stacking symmetry along the *c*-axis, individual monolayers are classified according to their in-plane coordination geometry: left- and right-tilted trigonal prisms (1H) are denoted as $1H_L$ and $1H_R$, and the corresponding left- and right-tilted octahedra (1T) as $1T_L$ and $1T_R$, respectively[16,17]. The 4H polytype exhibits a periodic stacking sequence of $1T_L$-$1H_L$-$1T_R$-$1H_R$, forming a hexagonal unit cell with the centrosymmetric space group $P6_3/mmc$ (No. 194). The individual trigonal-prismatic H layers retain local inversion asymmetry, allowing layer-dependent spin-orbit textures within the globally centrosymmetric crystal[43]. A larger-area STEM image (~ 50 × 50 nm$^2$) of this phase (Fig. S4) further confirms the periodic 1H/1T heterostructure distribution, consistent with the 4H phase assignment established by diffraction.

The atomic arrangements and structural parameters within the 1H/1T superlattice were determined by combined powder and single-crystal XRD (PXRD and SXRD). Rietveld refinement of the PXRD pattern (Fig. S3b) precisely determines the atomic coordinates, while the SXRD pattern (Fig. S3c) exhibits a well-defined reciprocal lattice free of extraneous reflections, confirming both high crystalline quality and phase purity[44,45]. Fig. 1(c) displays a representative cross-sectional, atomic resolution STEM image of the 6R-$NbSe_{2-x}Te_x$ ($x$ = 0.9) polytype acquired along the [110] zone axis. In this phase, adjacent 1H/1T bilayers stack with a systematic in-plane translation of (1/3$a$, -1/3$b$), generating a repeating sequence of $1H_L$-$1T_R$-$1H_L$-$1T_R$-$1H_L$-$1T_R$. This translated arrangement defines the six-layer repeat in the conventional hexagonal setting of space group $R3m$, and quantitative XRD refinements (Fig. S3e) confirm the six-layer superstructure, yielding lattice parameters $a = b = 3.524$ Å and $c = 39.243$ Å.

The Rietveld refinement yields modestly elevated weighted-profile $R$-factors ($R_{wp}$) of ~ 13% and ~ 12% for the 4H and 6R polytypes, respectively—an outcome primarily attributable to layer-stacking faults, a common feature in such intergrown vdW structures. Large-area STEM imaging of the 6R-phase crystal (Fig. S2, and S5) reveals localized deviations such as $1H_L$-$1H_L$ motifs within a predominantly ordered 6R sequence. This fault distribution is highly consistent with observations in related chalcogen-substituted transition-metal dichalcogenide systems. Importantly, these local stacking deviations introduce structural disorder not fully captured by the idealized refinement model, which broadens the PXRD peaks and increases residual electron density in SXRD measurements. The crystallographic parameters from diffraction refinements of the 4H, 6R and distorted 1T' specimens are summarized in Supplementary Tables S1-S3.

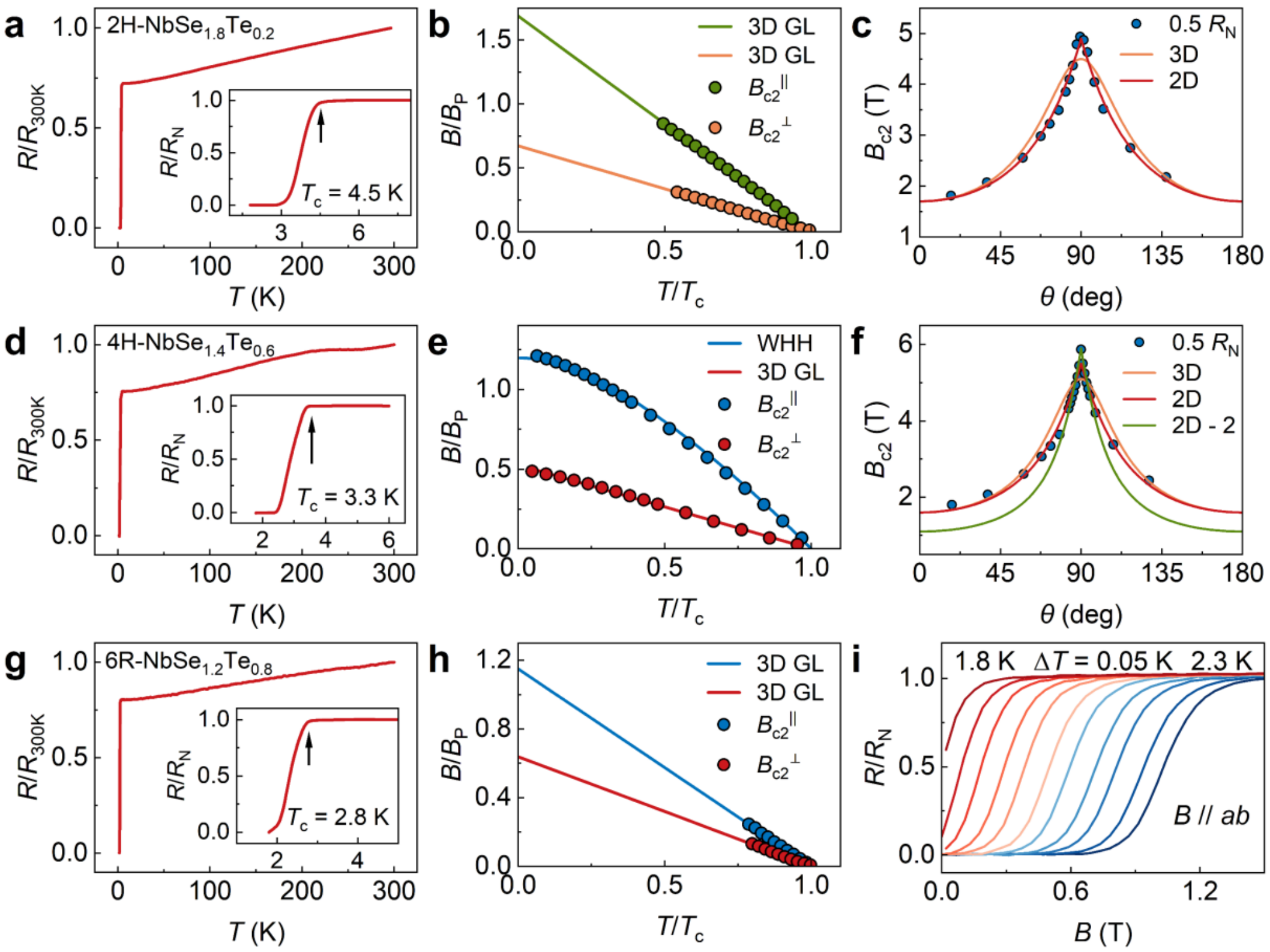

**Figure 2.** (a) Normalized resistance of 2H-$NbSe_{1.8}Te_{0.2}$ as a function of temperature; the inset provides a close-up of the low-temperature regime, where a superconducting transition is resolved at 4.5 K. (b) Upper critical field $B_{c2}$ versus temperature for in-plane (green) and out-of-plane (yellow) field orientations. The solid curves are fits to the 3D GL framework, describing the measured temperature dependence. (c) $B_{c2}$ anisotropy at 1.8 K, accurately reproduced by the 2D Tinkham model. (d) Resistive transition of 4H-$NbSe_{1.4}Te_{0.6}$, with a superconducting transition temperature of 3.3 K. (e) While $B_{c2}^{//}$ is analyzed within the WHH theory, $B_{c2}^{\perp}$ retains a conventional 3D GL form. (f) Angular-resolved $B_{c2}$ at 1.8 K (same $R_{xx}$ criterion); the enhanced $B_{c2}$ near in-plane directions is reproduced via two local 2D Tinkham descriptions of the broad cusp and the additional near-plane enhancement. (g) For the 6R-$NbSe_{1.2}Te_{0.8}$ phase, the superconducting transition shifts down to 2.8 K. (*h*) In this polytype, the upper critical field exhibits a linear temperature dependence, described by the anisotropic GL temperature dependence. (i) Isothermal resistance-versus-field sweeps for *B*//*ab* resolve the superconducting transition.

We next examine how progressive structural evolution governs the superconducting properties. At low tellurium content, the $NbSe_{2-x}Te_x$ system retains the pristine 2H structure. Fig. 2(a) shows the temperature-dependent normalized resistance for 2H-$NbSe_{1.8}Te_{0.2}$, which exhibits conventional metallic behavior down to a superconducting transition temperature $T_c$ of ~ 4.5 K. The residual-resistivity ratio is notably small ($R_{300K}/R_N < 1.5$, where $R_N = R_{xx}(10\ K)$), indicating substantial momentum scattering from chalcogen disorder and residual stacking faults. The upper critical fields $B_{c2}$ (where $B_{c2}$ is defined as the field at which the in-plane resistance $R_{xx}$ reaches 50% of its normal-state value) for both parallel ($B_{c2}^{//}$) and perpendicular ($B_{c2}^{\perp}$) magnetic fields exhibit a nearly linear temperature dependence, consistent with the 3D Ginzburg-Landau (GL) model, as seen in Fig. 2(b). Fitting yields GL length scales for the two field orientations of 4.85 nm for $B//ab$ and 7.69 nm for $B//c$, with the spatial components defined in Methods. Moreover, the angular dependence of $B_{c2}$ at 1.8 K displays a sharp cusp near the in-plane orientation (Fig. 2(c)), which is well described by the 2D Tinkham model rather than the anisotropic 3D GL theory[33,34]—a hallmark of the quasi-2D nature of the 2H phase.

With increasing Te substitution, the system transforms into the 4H polytype, and the superconducting response undergoes a qualitative change. The 4H-$NbSe_{1.4}Te_{0.6}$ crystal exhibits a slightly reduced $T_c$ of 3.3 K, yet its in-plane upper critical field $B_{c2}^{//}$ develops pronounced non-linear upward curvature (Fig. 2(e)), contrasting with the approximately linear response of the 2H specimen. The small normal-state resistance ratio $R(300\ K)/R(5\ K) \approx 1.4$ indicates a substantial residual-scattering contribution. Together with substitutional Se/Te disorder and the local stacking defects resolved in the series, it supports the dirty-limit WHH treatment. The WHH description captures this enhancement through the interplay of orbital and paramagnetic depairing with spin-orbit scattering[35,38]. The low-temperature parallel field exceeds the weak-coupling Pauli reference $B_P \approx 1.84T_c$, while the near-plane angular enhancement represents a separate feature of the superconducting response.

The dimensionality of the 4H phase is further distinguished by the angular dependence of $B_{c2}(\theta)$. For 4H-$NbSe_{1.4}Te_{0.6}$, an additional enhancement appears within a narrow angular window near the *ab* plane (Fig. 2(f)). Two local Tinkham descriptions resolve the broad angular cusp and the extra near-plane enhancement: the outer-angle fit gives $B_{c2}^{//}$ = 5.1 T and $B_{c2}^{\perp}$ = 1.6 T, whereas the near-plane fit gives $B_{c2}^{//}$ = 5.9 T and $B_{c2}^{\perp}$ = 1.1 T. The second 4H crystal reproduces this characteristic angular feature (Fig. S13). Similar narrow-angle enhancements occur in the bulk superlattices $Ba_6Nb_{11}S_{28}$ and $BaTa_2S_5$[23,46]. In 4H-$NbSe_{2-x}Te_x$, the distinct H/T stacking and local layer symmetry provide a natural setting for spin-orbit-assisted enhancement together with reduced interlayer coherence[23,43,47-51].

For the rhombohedral 6R phase ($NbSe_{1.2}Te_{0.8}$), the transition temperature decreases to 2.8 K and both field orientations follow an approximately linear GL temperature dependence. The corresponding GL length scales for the two field orientations are 7.45 nm for *B*//*ab* and 10.00 nm for *B*//*c*. Together with its smaller critical-field anisotropy, this response indicates greater coherence between superconducting blocks than in 4H. The translated 6R stacking changes the orbital-overlap pathways between H/T blocks, giving a more three-dimensional superconducting response even though the crystal remains anisotropic.

At the highest Te contents, the $NbSe_{2-x}Te_x$ layers preferentially assemble into T-derived configurations. The T-derived specimens in our series exhibit lower transition temperatures than the H-type and mixed H/T specimens (Fig. 5). Their lower transition temperatures differ from the approximately 3.2 K reported for 1T-NbSeTe single crystals[22].

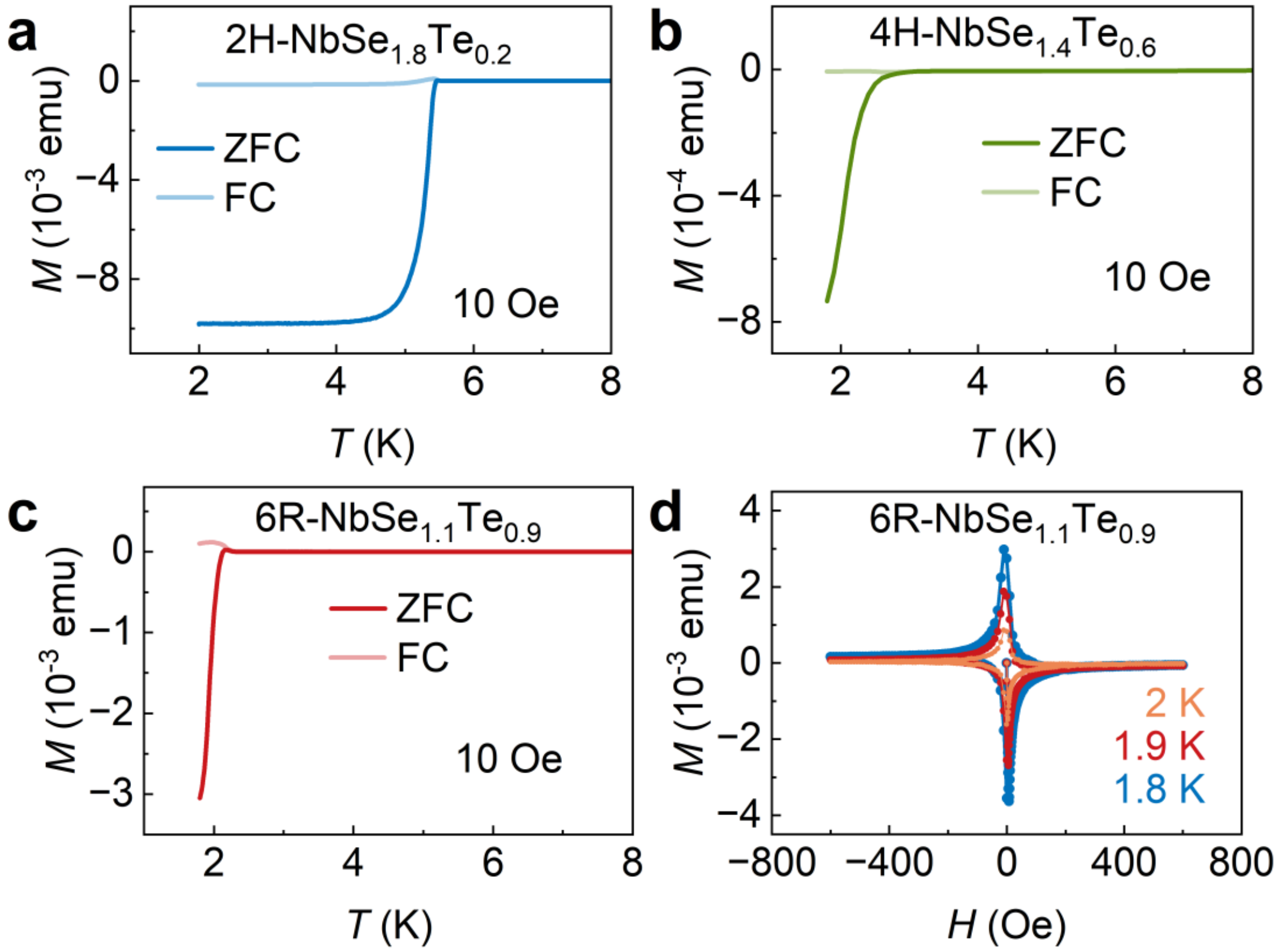

**Figure 3.** (a-c) Temperature dependencies of dc magnetic moment for 2H-$NbSe_{1.8}Te_{0.2}$, 4H-$NbSe_{1.4}Te_{0.6}$ and 6R-$NbSe_{1.1}Te_{0.9}$, respectively. The superconducting transition in each sample is clearly confirmed by the onset of a diamagnetic shielding response in the temperature-dependent magnetization measurements. (d) Magnetic hysteresis loops of 6R-$NbSe_{1.1}Te_{0.9}$ recorded at various temperatures under an out-of-plane magnetic field. The loops exhibit characteristic superconducting diamagnetic behavior, with the magnitude of hysteresis decreasing with increasing temperature, which is consistent with reduced vortex pinning strength near the superconducting transition temperature $T_c$.

The superconducting response across these structural phases is corroborated by dc magnetization measurements. Temperature-dependent magnetic-moment curves for the 2H, 4H, and 6R samples reveal strong diamagnetic shielding with a clear separation between zero-field-cooled (ZFC) and field-cooled (FC) traces—characteristic of robust vortex pinning in layered superconductors (Fig. 3(a-c)). Magnetic hysteresis loops recorded under out-of-plane fields confirm this diamagnetic behavior, with the loop width progressively diminishing as the temperature approaches $T_c$ (Fig. 3(d)). The specific-heat anomaly of 6R-$NbSe_{1.2}Te_{0.8}$ provides complementary thermodynamic evidence for bulk superconductivity in that specimen (Fig. S17).

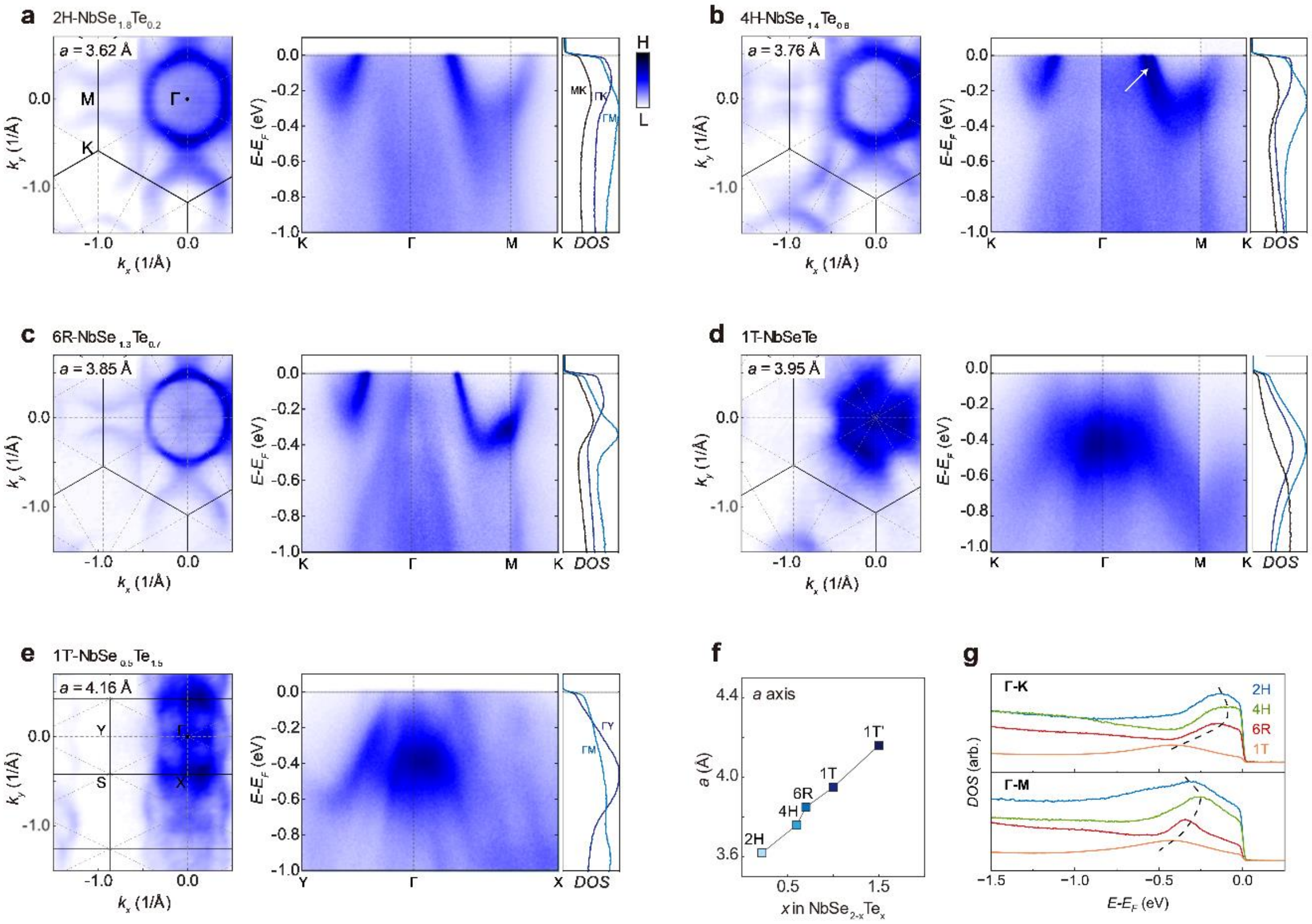

**Figure 4.** (a-e) Fermi-surface maps (left), band dispersions (middle) and integrated ARPES intensities (right) for 2H-$NbSe_{1.8}Te_{0.2}$, 4H-$NbSe_{1.4}Te_{0.6}$, 6R-$NbSe_{1.3}Te_{0.7}$, 1T-NbSeTe and 1T'-$NbSe_{0.5}Te_{1.5}$, respectively. The corresponding band dispersions and integrated spectra are shown along Γ-K, Γ-M and K-M; the 1T' spectra are collected along Γ-X and Γ-Y. (f) In-plane lattice estimates inferred from the apparent reciprocal-space dimensions in the ARPES maps. (g) Comparison of integrated ARPES intensities along Γ-K and Γ-M. These selected-cut spectra are the quantities labelled DOS in the panels.

To elucidate the electronic reconstruction accompanying the polytypic transitions, we performed ARPES to track the systematically evolving electronic structures across the $NbSe_{2-x}Te_x$ system (Fig. 4). The left panels of Fig. 4 display the characteristic Fermi-surface topologies for representative Te substitution levels. For the 2H ($x$ = 0.2), 4H ($x$ = 0.6), and 6R ($x$ = 0.7) phases, the Fermi surfaces consistently exhibit distinct hexagonal geometries with prominent hole pockets centered at the Γ and K high-symmetry points—a hallmark electronic signature of the H-type layers[6,7,16,20]. By evaluating the dimensions of the reconstructed Brillouin zones extracted from these maps, we systematically infer the in-plane lattice constant evolution (Fig. 4(f)), which follows the same overall expansion trend as room-temperature XRD in Fig. 5(a). The numerical estimates differ because the ARPES construction uses visible electronic contours, whose orbital-dependent intensity and surface sensitivity differ from the bulk crystallographic periodicity measured by XRD. In sharp contrast, the Fermi surface of the monoclinic 1T' phase ($NbSe_{0.5}Te_{1.5}$) transforms into a highly anisotropic, chain-like configuration with substantially reduced symmetry, directly reflecting the underlying monoclinic lattice distortion.

Beyond the Fermi-surface geometry, the energy dispersions reveal the electronic reconstruction associated with the H/T stacking. A split feature is visible along Γ-M in the 4H spectrum (white arrow in Fig. 4(b)). SOC calculations distinguish the momentum and symmetry dependence of this reconstruction (Figs. S19-S22): the centrosymmetric 4H model retains double degeneracy, while the non-centrosymmetric 6R model shows an Ising-type splitting of approximately 0.185 eV near K/H. No obvious near-$E_F$ spin splitting appears along Γ-M/A-L in the virtual-crystal bands. The observed Γ-M branches are consistent with a reconstructed, H/T-hybridized multiband spectrum

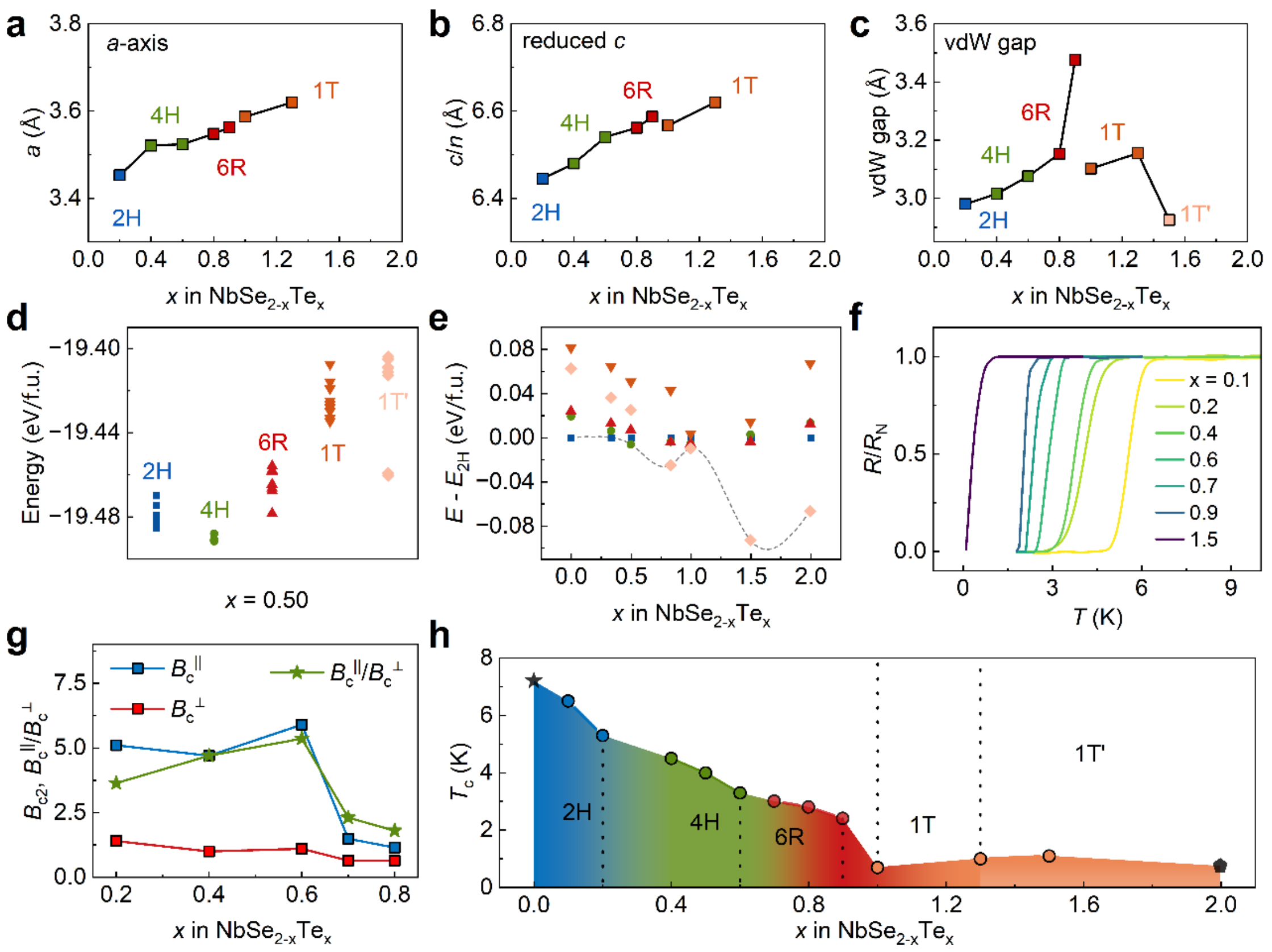

**Figure 5.** (a) In-plane lattice parameter $a$; (b) reduced $c$-axis parameter $c/n$; and (c) vdW gap thickness, defined as the vertical separation between chalcogen atoms across adjacent layers. Data for the 2H, 4H, 6R, and 1T phases are marked by blue, green, red, and yellow squares, respectively. (d) Calculated formation energies for representative 2H, 4H, 6R, 1T, and distorted 1T' configurations at $x = 0.50$; multiple points within each structural type correspond to inequivalent Se/Te substitution arrangements. (e) Relative formation energies of the competing polytypes as a function of $x$, referenced to the 2H phase at each Te concentration. The dashed line traces the lowest calculated energy among the sampled configurations, serving as a guide to the phase competition. (f) Temperature-dependent normalized resistance ($R/R_N$) showing the evolution of the superconducting transition with Te substitution. (g) Estimated upper critical fields $B_{c2}$ for fields applied parallel (blue, $B_{c2}^{//}$) and perpendicular (red, $B_{c2}^{\perp}$) to the $ab$-plane. Across the 2H-to-4H crossover ($0.2 < x < 0.6$), $B_{c2}$ exhibits little variation despite the decreasing $T_c$; in contrast, the anisotropy ratio $B_{c2}^{//}/B_{c2}^{\perp}$ (green stars) increases steadily—consistent with reduced effective interlayer coherence in the 4H matrix. The anisotropy ratio is dimensionless. (h) Comprehensive phase diagram summarizing the compositional stability ranges of the 2H,

4H, 6R, 1T, and distorted 1T' phases, alongside the overall suppression of $T_c$ with increasing $x$, including the small recovery at high Te content. The nominal regimes progress from 2H through mixed-layer 4H and 6R to ideal 1T and distorted 1T' extending to $NbTe_2$ at $x = 2$. Notably, the steepest drop in $T_c$ occurs at the 6R-to-1T structural boundary, highlighting the correlation between the structural reconstruction and the superconducting response.

and are distinct from the calculated K/H Ising splitting. Local H-layer SOC remains compatible with the global inversion symmetry of 4H[43]. The integrated spectra in Fig. 4(g) show a redistribution of occupied spectral weight as the H/T coordination changes. Their intensities reflect the electronic states together with photoemission matrix elements, orbital character and surface sensitivity. The combined ARPES and calculated dispersions link the structural evolution to the electronic reconstruction accompanying the superconducting phase diagram.

To establish a comprehensive structural and electronic overview of the $NbSe_{2-x}Te_x$ system, we construct a unified phase diagram incorporating both lattice parameters and superconducting characteristics (Fig. 5). Fig. 5(a) shows the room-temperature in-plane lattice parameter $a$ determined from diffraction; the local STEM profiles are discussed separately in SI Section IV. As the larger Te atoms substitute for Se across the entire composition range, the average in-plane lattice parameter increases with increasing $x$. To systematically evaluate the structural stability of the various stacking configurations, we normalize the $c$-axis lattice parameter by the number of layers ($n$) within each stacking repeat unit. As shown in Fig. 5(b), this reduced parameter ($c/n$) increases smoothly in the low-to-intermediate substitution regime, but exhibits a distinct discontinuity at the 6R-1T phase boundary. This prominent structural jump signals a phase reconstruction, consistent with the calculated formation-energy landscape (Fig. 5(e))[25-27].

Another critical parameter dictating the physical properties is the vdW gap thickness, quantitatively extracted from $c/n$ by subtracting the vertical separation between adjacent chalcogen layers—derived from the refined $z$-coordinates of the chalcogen atoms above and below the transition-metal plane. As presented in Fig. 5(c), the vdW gap initially expands rapidly with increasing Te content through the 4H and 6R phases. Crucially, a significant collapse occurs precisely at the 6R-to-1T structural transition. The non-monotonic separation accompanies the changes in coordination and stacking registry[16-18,21,50]. Effective interlayer coupling also depends on lateral registry and orbital overlap, so it need not vary monotonically with the geometric vdW gap.

Formation-energy calculations provide further insight into the phase competition (Figs. 5(d,e) and Fig. S18). At each composition, inequivalent Se/Te configurations give an energy distribution for each polytype. Te substitution reduces the energy separation among the competing H/T and T-derived structures. The close competition makes local alloy environments and structural reconstruction important. The plotted energies describe unreconstructed polytype cells; published calculations show that a CDW reconstruction in the T layers can lower the energy of a related H/T superlattice[16]. Configurational sampling and retention of closely competing structures during high-temperature CVT growth followed by quenching further connect the calculated landscape to the observed phase sequence. The sampled energies provide qualitative context for phase competition rather than quantitative equilibrium phase boundaries.

We now turn to the macroscopic evolution of superconductivity. The temperature-dependent normalized resistance ($R/R_N$) curves across the entire substitution series (Fig. 5(f)) show that while the monoclinic 1T' phase at $x = 1.5$ exhibits a severely suppressed transition below 1 K, all self-assembled 1H/1T superlattice polytypes maintain robust transitions with $T_c > 2$ K. Fig. 5(g) compares the estimated upper critical fields and the resulting anisotropy parameters ($B_{c2}^{//}/B_{c2}^{\perp}$) as a function of $x$. Across the 2H-to-4H crossover ($0.2 < x < 0.6$), $B_{c2}$ shows no dramatic variation despite the gradual decrease in $T_c$; however, the anisotropy ratio increases steadily, reaching its maximum in the 4H phase. This enhancement, together with the angular cusp and the narrow near-plane feature, supports reduced effective interlayer coherence and enhanced low-dimensional superconductivity in 4H[23,47-50].

Normal-state transport provides a complementary comparison of the layer anisotropy (Fig. S23). For representative 2H-$NbSe_{1.8}Te_{0.2}$, 4H-$NbSe_{1.6}Te_{0.4}$ and 6R-$NbSe_{1.1}Te_{0.9}$ crystals, $\rho_{zz}/\rho_{xx}$ is enhanced toward low temperature and reaches its largest values in 4H. The increase from 2H to 4H and decrease toward 6R agree with the evolution of superconducting two-dimensionality. This correspondence supports the role of stacking-dependent interlayer charge motion in the superconducting anisotropy.

In sharp contrast, the rhombohedral 6R phase ($x \approx 0.6$–$0.9$) exhibits an abrupt reduction of both $B_{c2}$ and the superconducting anisotropy, signaling a crossover back toward more three-dimensional superconductivity. This non-monotonic evolution correlates with the polytype sequence, while Te content, interlayer spacing, electronic structure and disorder vary concurrently. The 4H polytype provides a stacking configuration that is compatible with reduced interlayer coherence and enhanced two-dimensional superconductivity, whereas the translated stacking sequence in the 6R phase appears to support stronger effective coupling between neighboring superconducting blocks. Thus, the emergence and disappearance of enhanced $B_{c2}$ anisotropy reflect a delicate balance between local polymorphic building blocks, long-range stacking order, and interlayer electronic hybridization. The present series does not isolate the contribution of stacking from concurrent chemical and structural changes; disorder and growth-related metastability can also contribute. The SOC band calculations further show that spin splitting and superconducting dimensionality are distinct: 6R has a finite Ising-type splitting near K/H while exhibiting the less anisotropic superconducting response. The stacking-dependent balance of local SOC and interlayer hybridization is therefore central to the 4H–6R comparison.

The overall structural and electronic landscape is summarized in the comprehensive phase diagram of Fig. 5(h). The pristine 2H phase transitions to the 4H polytype for $x \approx 0.4$–$0.6$, where enhanced 2D superconductivity is stabilized. With further Te substitution, the 4H structure becomes unstable, giving way to the rhombohedral 6R polytype ($x \approx 0.6$–$0.9$), which supports a more three-dimensional superconducting response relative to 4H. Above $x \sim 1.0$, pure T-coordinated configurations emerge, with the ideal 1T phase persisting until $x = 1.3$, followed by the high-Te distorted 1T' regime, including the monoclinic $NbTe_2$ endpoint[42]. Across the phase diagram, $T_c$ is suppressed overall with Te incorporation, with a small recovery in the high-Te regime; the most precipitous drop occurs at the 6R-to-1T structural reconstruction—linking the structural reconstruction to concurrent changes in the electronic states and superconducting dimensionality.

**Conclusion**

We have systematically investigated the tellurium-substitution-driven structural phase transitions in $NbSe_{2-x}Te_x$ single crystals and elucidated their profound impact on superconducting dimensionality. By combining complementary characterization techniques, we have established a comprehensive phase diagram of the measured series spanning the entire substitution range. The low-Te members retain the 2H structure; intermediate substitution triggers spontaneous 1H/1T heterostructuring, sequentially stabilizing the 4H ($x \approx 0.4 - 0.6$) and 6R ($x \approx 0.6 - 0.9$) polytypes; high Te content ($x > 0.9$) drives a transition into pure T-coordinated configurations, culminating in the distorted monoclinic 1T' phase at the highest doping levels. The combined transport, ARPES and SOC calculations connect this evolution of stacking sequence and space group to changes in superconducting anisotropy and electronic band structure.

Each polytype exhibits a distinct superconducting fingerprint. The 2H phase ($NbSe_{1.8}Te_{0.2}$) retains a metallic normal state with moderate 2D character. In the 4H regime, the superconducting anisotropy is markedly enhanced: the non-linear upward curvature in $B_{c2}^{//}$ and the anomalous in-plane $B_{c2}$ enhancement, exceeding conventional 2D Ginzburg-Landau predictions, support a stacking-dependent reduction of effective interlayer coherence together with spin-orbit-assisted enhancement. In stark contrast, the 6R phase ($NbSe_{1.2}Te_{0.8}$) exhibits a reduced $T_c$ (~ 2.8 K) and a reentrant linear $B_{c2}^{//}(T)$ behavior, consistent with greater coherence between superconducting blocks and a more three-dimensional response relative to 4H. The T-derived specimens at high Te content have lower transition temperatures in this series, differing from the approximately 3.2 K reported for 1T-NbSeTe[22].

These results establish that the superconducting evolution of self-assembled 1H/1T superlattices reflects the combined effects of interlayer electronic coherence and symmetry-dependent spin-orbit protection. The antiparallel, centrosymmetric 4H stack and the translated, non-centrosymmetric 6R stack provide different interlayer-hybridization pathways and local spin-orbit environments, even though both contain H/T building blocks. Notably, the 4H phase appears to define an optimal regime in which ordered 1H/1T stacking and Te-induced changes in the local electronic structure jointly favor enhanced anisotropic superconductivity. This work thus deepens the fundamental understanding of phase engineering in van der Waals materials and provides a concrete design principle, tailoring layer-stacking symmetry to control superconducting dimensionality, for the development of TMD-based quantum devices.

## ASSOCIATED CONTENT

### Supporting Information

The Supporting Information includes EDS characterizations, atomic-scale structural analyses, diffraction refinements, magnetotransport, directional normal-state resistivity and specific-heat measurements, configurational-energy calculations, and SOC band structures of $NbSe_{2-x}Te_x$ (Figs. S1-S23 and Tables S1-S3).

## AUTHOR INFORMATION

### Corresponding Author

chencheng1@shanghaitech.edu.cn; zzyang@phy.ecnu.edu.cn; kliu@ruc.edu.cn; guoyf@shanghaitech.edu.cn.

### Author Contributions

† X.Q.L., H.N.W., Z.G.L. and P.H.S. contributed equally to this work.

### Notes

The authors declare no competing financial interest.

## ACKNOWLEDGMENTS

The authors acknowledge the National Key R&D Program of China (Grants No. 2023YFA1406100, 2024YFA1408400 and 2022YFA1403103). Y.F.G. acknowledges the Science and Technology Commission of Shanghai Municipality (25DZ3008200) and the support by the Key Laboratory of Aeronautical Materials and Application Technology (ZHKF-250103). Z.Z.Y. acknowledges financial support from Science and Technology Commission of Shanghai Municipality (Grant No. 24JD1401200). W.X. thanks the support by the Shanghai Sailing Program (23YF1426900). C.C. thanks the financial support from the National Key Research and Development Program of China (grant no. 2024YFA1410400). P.H.S acknowledges the supports by the National Natural Science Foundation of China (Grant No. 12404266) and the Science Research Project of Hebei Education Department (Grant No. BJ2025060). The authors also thank the support from Analytical Instrumentation Center (#SPST-AIC10112914) and the Double First-Class Initiative Fund of ShanghaiTech University. Computational resources have been provided by the Physical Laboratory of High-Performance Computing at Renmin University of China and the Hefei advanced computing center.

BRIEFS: Composition-driven polytypism in $NbSe_{2-x}Te_x$ links H/T stacking to distinct superconducting responses. The centrosymmetric antiparallel 4H superlattice exhibits enhanced critical-field anisotropy and a narrow near-plane enhancement, whereas translated 6R stacking supports a more three-dimensional response.

**TOC:**

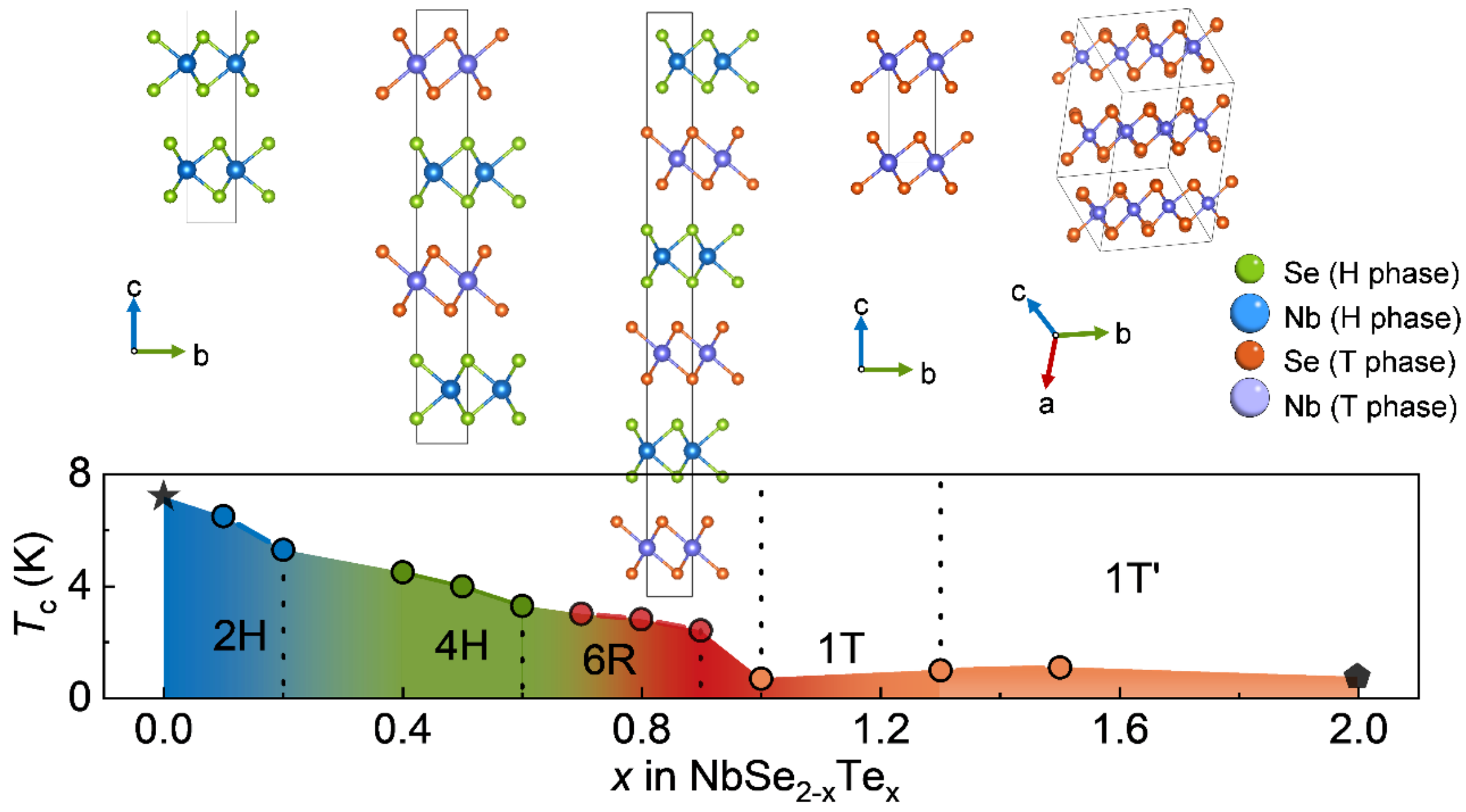

c
b
c
b
c
b
a
Se (H phase)
Nb (H phase)
Se (T phase)
Nb (T phase)
8
4
0
$T_c$ (K)
2H
4H
6R
1T
1T'
0.0
0.4
0.8
1.2
1.6
2.0
$x$ in $NbSe_{2-x}Te_x$

## I. Energy dispersive spectroscopy (EDS) characterizations of $NbSe_{2-x}Te_x$ polytypes

Fig. S1 presents representative crystal morphologies and EDS results for the 2H, 4H, 6R, 1T and distorted 1T' members. EDS identifies Nb, Se and Te and tracks the change in Se/Te composition across the series, with a compositional uncertainty within 10%. Together with the mixed Se/Te diffraction sites and atomic-resolution STEM, these results support Te substitution on the Se/Te chalcogen sublattice[1-4]. The precursor molar ratio specifies the growth mixture. Nb occupancy was fixed at 1.0 in all structural refinements; allowing it to vary yielded values above unity for the adopted Nb sites.

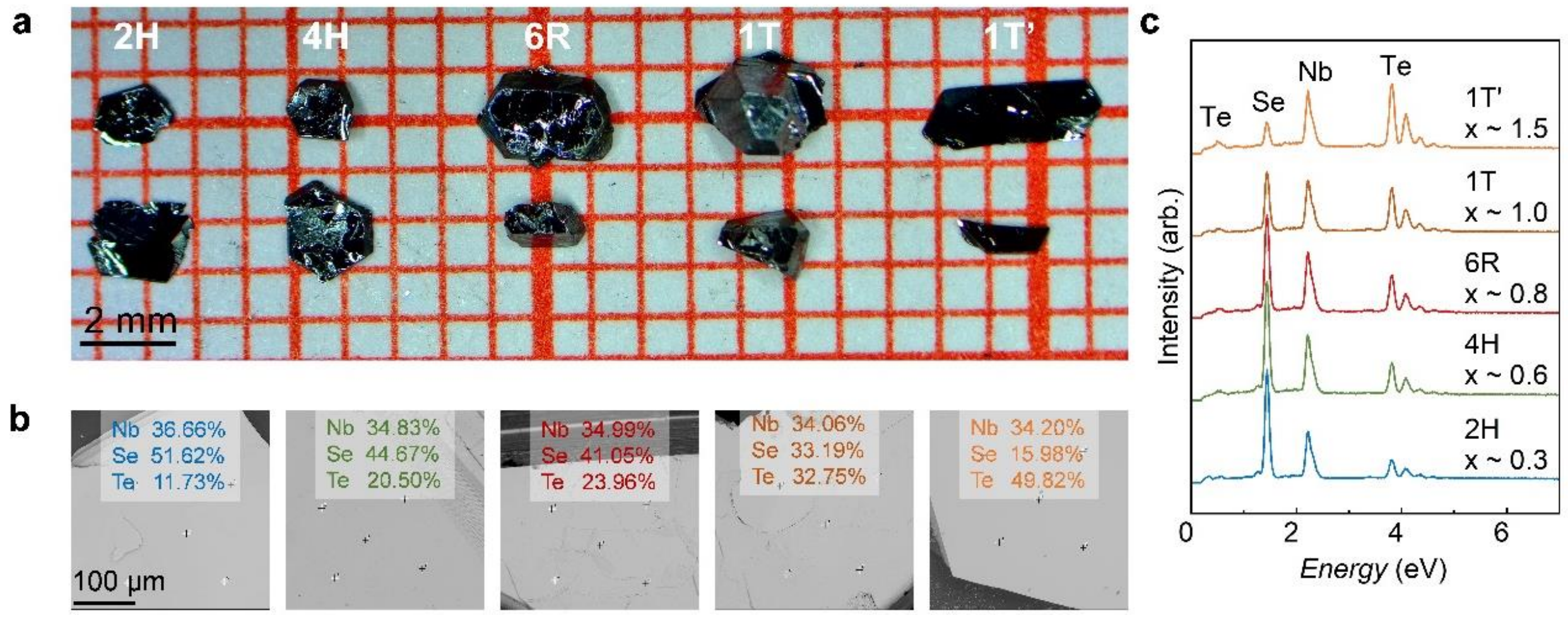

**Figure S1.** (a) Optical microscopy images of $NbSe_{2-x}Te_x$ with varying Te doping levels, corresponding to the 2H, 4H, 6R, 1T, and distorted 1T' polytypic phases, respectively. (b) EDS elemental-composition summary for representative $NbSe_{2-x}Te_x$ crystals. (c) Corresponding EDS spectrum, where the Te characteristic peak exhibits a pronounced enhancement relative to the Nb signal, confirming successful Te incorporation. Elemental percentages in panel (b) are atomic percentages.

The cross-sectional EDS maps in Fig. S2 show Se and Te distributed throughout the imaged 6R H/T stack, with no resolved Se-rich or Te-rich segregated domain. The resolved Nb/chalcogen columns in STEM and the mixed Se/Te-site model used in the SXRD and PXRD refinements consistently support substitution on the chalcogen sublattice[2,3]. These observations establish the spatial distribution across the observed mixed-layer structure. The formation-energy calculations sample inequivalent Se/Te arrangements and show that Te substitution reduces the energy separation between competing H- and T-derived stackings, providing an energetic basis for their assembly into mixed-layer structures.

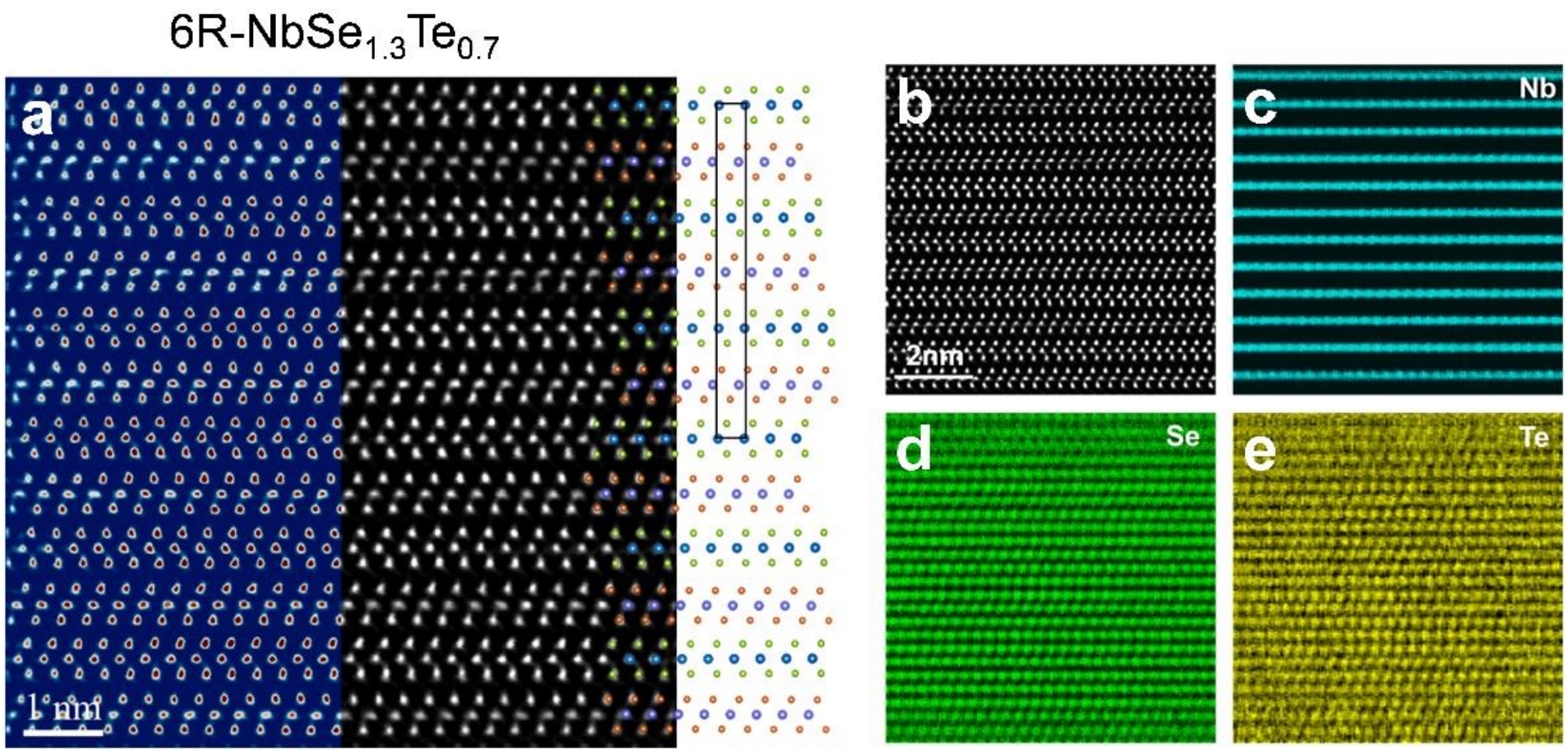

**Figure S2.** (a) HAADF-STEM image of a cross-sectional 6R-$NbSe_{1.3}Te_{0.7}$ single crystal. The alternating 1H/1T heterostructure stacking sequence is overlaid with the corresponding 6R atomic structural model. (b-e) Atomic-resolution EDS elemental maps acquired from the same region.

## II. Structural analysis of the $NbSe_{2-x}Te_x$ single crystals with 1H/1T superlattice

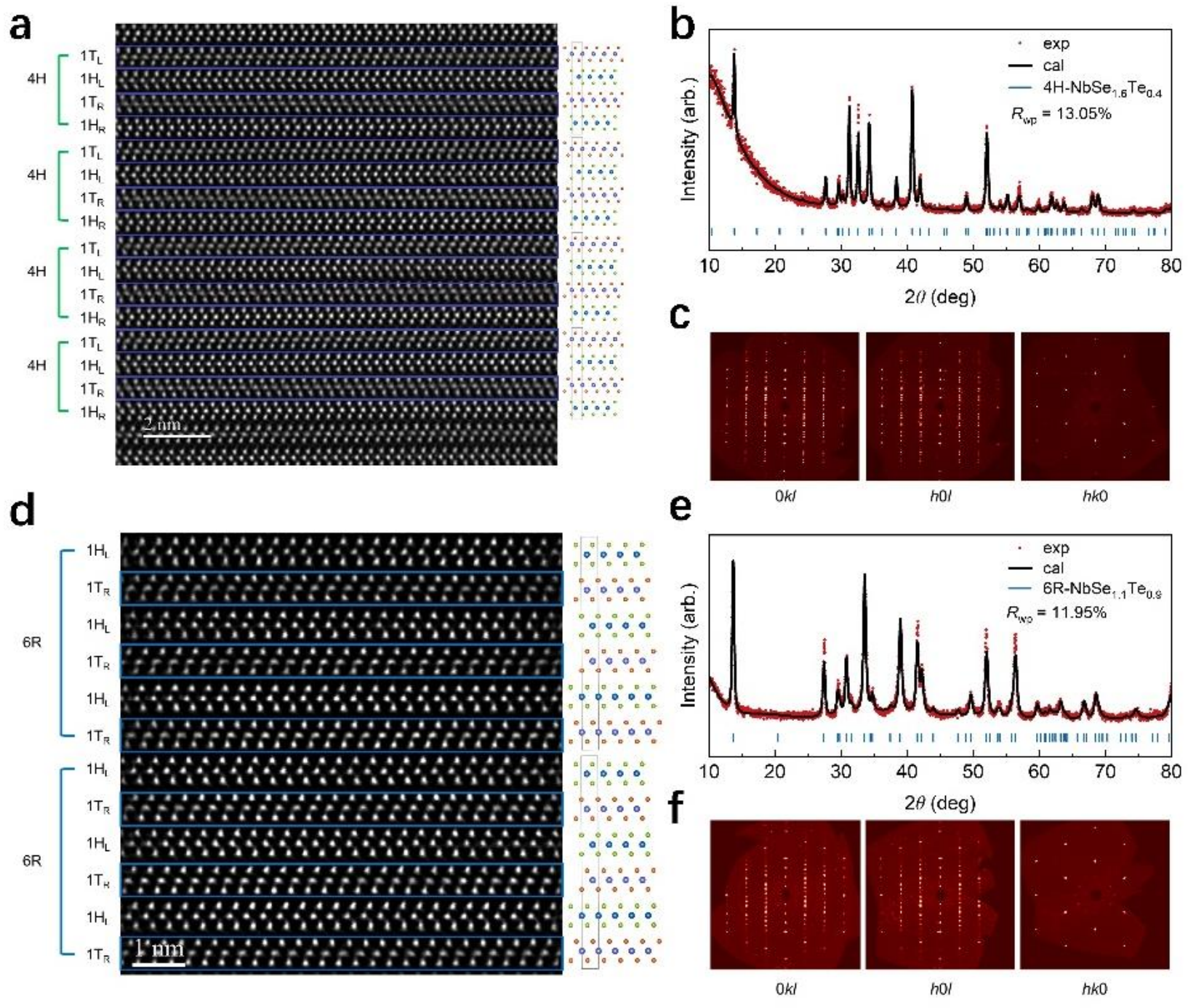

**Figure S3.** (a) Cross-sectional STEM image of the 4H-type $NbSe_{1.6}Te_{0.4}$ superlattice viewed along the [110] zone axis. The 1H and 1T layers are distinguished by their in-plane symmetry, namely, the tilt direction of trigonal prisms ($1H_L/1H_R$) and octahedra ($1T_L/1T_R$), and the alternating stacking sequence ($1T_L$-$1H_L$-$1T_R$-$1H_R$) is overlaid with the 4H atomic structural model. (b) Rietveld refinement of the PXRD pattern, and (c) corresponding SXRD patterns reconstructed from the (*0kl*), (*h0l*), and (*hk0*) reciprocal planes for $NbSe_{1.6}Te_{0.4}$. (d) Cross-sectional STEM image of the 6R-type $NbSe_{1.1}Te_{0.9}$ superlattice along [110], with the $1H_L$-$1T_R$-$1H_L$-$1T_R$-$1H_L$-$1T_R$ stacking sequence overlaid on the 6R structural model. (e) Rietveld refinement of the PXRD pattern, and (f) SXRD patterns along the same reciprocal-space planes for $NbSe_{1.1}Te_{0.9}$.

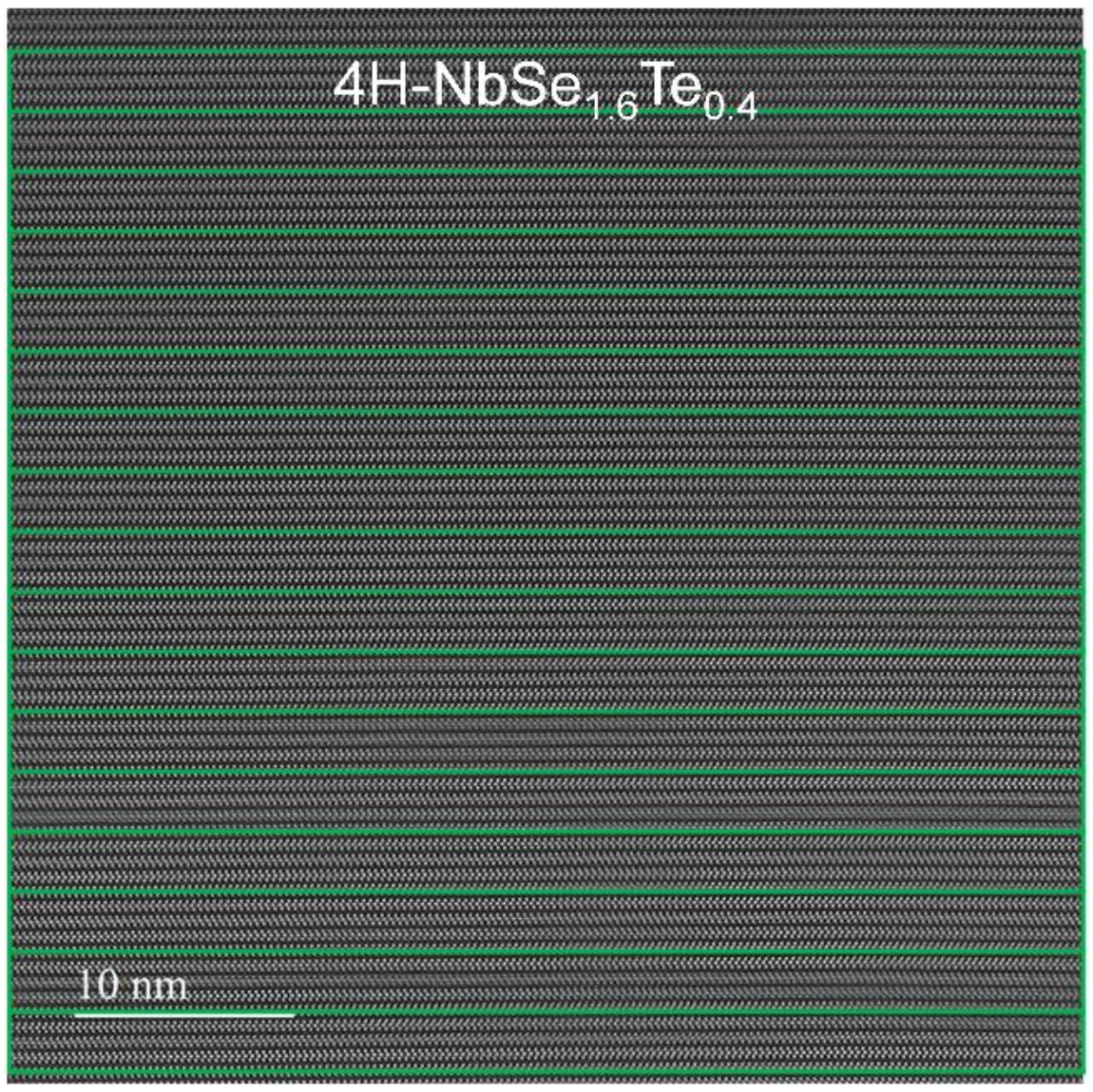

**Figure S4.** Cross-sectional overview of 4H-$NbSe_{1.6}Te_{0.4}$, showing homogeneous contrast, periodic 4H unit-cell stacking (green squares), and the absence of line defects or phase segregation within the observed region.

In 1H/1T superlattices, the random intermixing of H-type and T-type layers introduces stacking faults that break translational symmetry along the crystallographic *c*-axis and may give rise to mixed-phase regions[2-4]. Nevertheless, a large-field STEM image (~ 50 × 50 $nm^2$) of 4H-$NbSe_{1.6}Te_{0.4}$ (Figs. S3-4) confirms the periodic ordering of 1H/1T heterostructures. The ideal 4H primitive unit cells are stacked periodically along the *c*-axis (highlighted by green squares), providing direct evidence for local 4H order, consistent with the phase assignment from PXRD and SXRD. The configurational-energy results provide a physical setting for this ordered mixed-layer stacking in the intermediate Te regime[2].

Upon further Te substitution, the six-layer 6R sequence becomes stabilized. The overview image of 6R-$NbSe_{1.1}Te_{0.9}$ (~25 × 25 $nm^2$, Fig. S5) shows localized departures, including adjacent $1H_L$-$1H_L$ motifs, within a predominantly alternating H/T stack. The STEM images identify the local fault motifs, while PXRD and SXRD establish the average 6R phase. This combination distinguishes local stacking disorder from the sample-scale structural assignment[3,4].

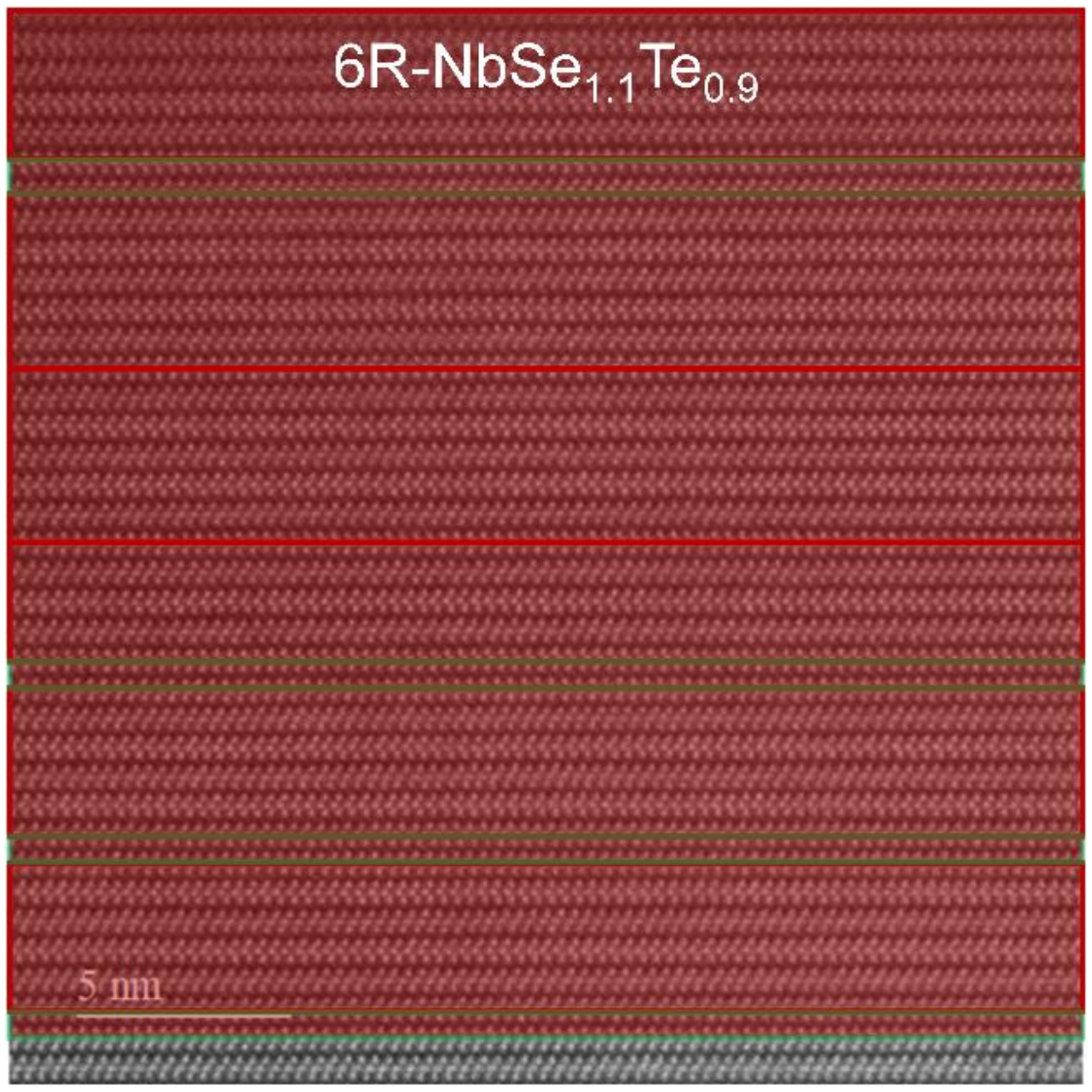

**Figure S5.** Cross-sectional overview image of 6R-$NbSe_{1.1}Te_{0.9}$. Red squares mark the six-layer 6R repeats in the conventional hexagonal setting, and the green squares mark the stacking fault (e.g., adopting a $1H_L$-$1H_L$ stacking motif).

## III. Structural characterization for distorted 1T'-$NbSe_{0.5}Te_{1.5}$ single crystals

At higher Te doping levels ($x > 0.9$), $NbSe_{2-x}Te_x$ initially adopts an ideal 1T phase (space group $P$-3$m$1, No. 164), which remains stable up to $x = 1.3$. Beyond this threshold, the system undergoes a further phase transition to a distorted 1T polymorph, denoted as 1T' (space group $C2/m$, No. 12)[1-6]. In this $C2/m$ structure, the distorted octahedral coordination around Nb atoms drives a unit-cell expansion, yet the overall arrangement retains centrosymmetry.

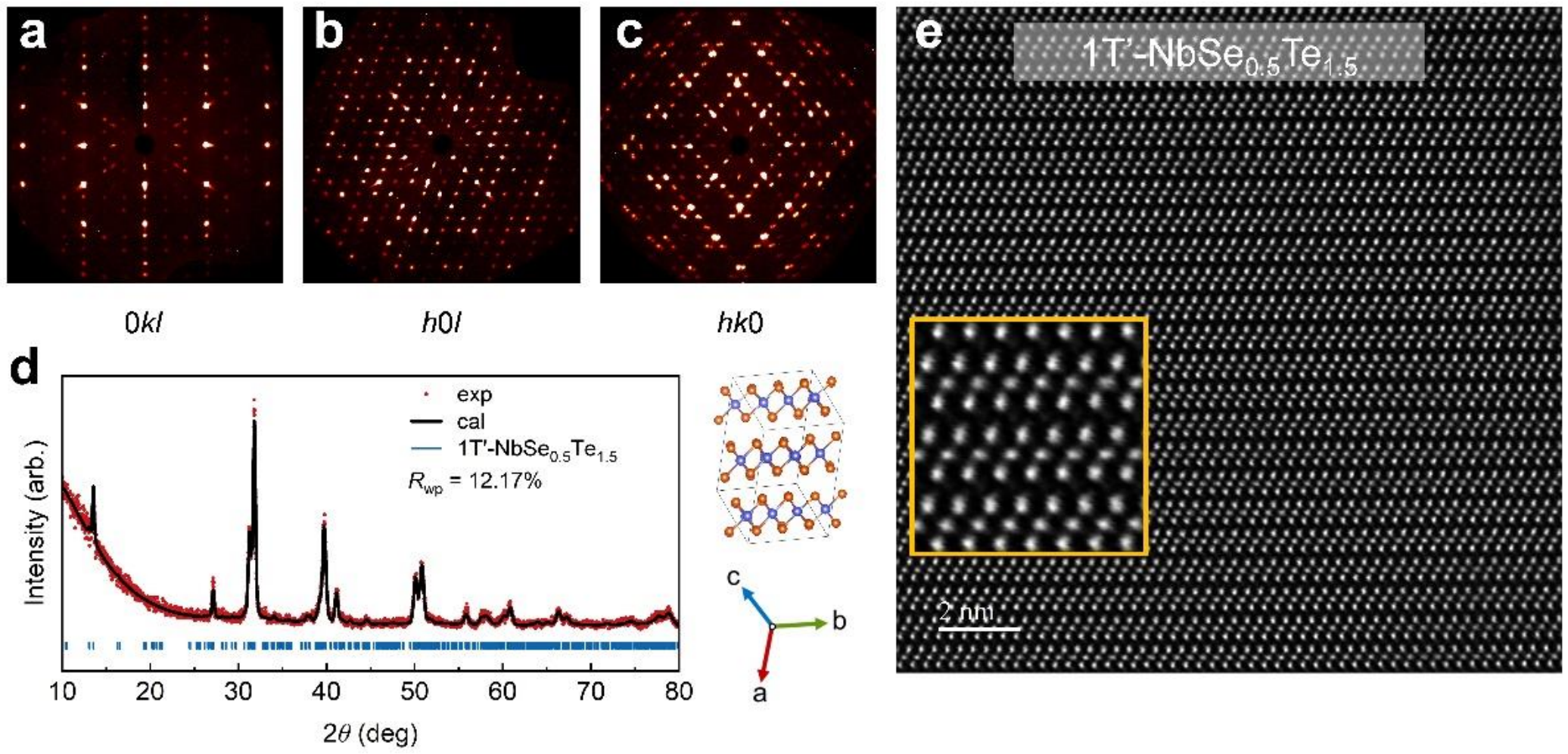

**Figure S6.** (a-c) SXRD patterns collected in reciprocal space along the (*0kl*), (*h0l*), and (*hk0*) planes for 1T'-$NbSe_{0.5}Te_{1.5}$. (d) Rietveld refinement results of the PXRD patterns. (e) Cross-sectional STEM image of the distorted 1T'-$NbSe_{0.5}Te_{1.5}$ single crystal with enlarged area.

Figs. S6 (a-c) present SXRD patterns of the distorted 1T'-$NbSe_{0.5}Te_{1.5}$ phase. Notably, the pattern from the *h0l* lattice plane unambiguously reveals monoclinic symmetry, which stands in clear contrast to the trigonal space group of the ideal 1T phase—thus providing direct crystallographic evidence for the structural distortion. Fig. S6d shows the PXRD pattern of the same composition. Rietveld refinement of these data enabled precise determination of atomic coordinates, further corroborating the monoclinic $C2/m$ structure.

Fig. S6e displays a cross-sectional STEM image of the distorted 1T' structure. When viewed along this zone axis, however, distinguishing the ideal 1T phase from its distorted 1T' counterpart proves challenging from atomic-scale cross-sectional images alone: the magnified region in Fig. S6e closely resembles the [110] projection of the 1T phase, underscoring the subtlety of the octahedral distortion that defines the 1T' polymorph.

## IV. Structural analysis of the $NbSe_{2-x}Te_x$ polytypes through STEM

To compare the stacking geometries, we measured the vdW gap and projected Nb-column separations in atomic-resolution STEM images. Figs. S7-S9 show 4H-$NbSe_{1.6}Te_{0.4}$, 6R-$NbSe_{1.3}Te_{0.7}$ and distorted 1T'-$NbSe_{0.5}Te_{1.5}$. The vdW gap is the vertical separation between chalcogen planes of neighboring layers. The in-plane profile measures the separation of projected Nb columns along the selected line. This projected separation is distinct from the crystallographic lattice parameter *a* and from a three-dimensional Nb-Nb bond length.

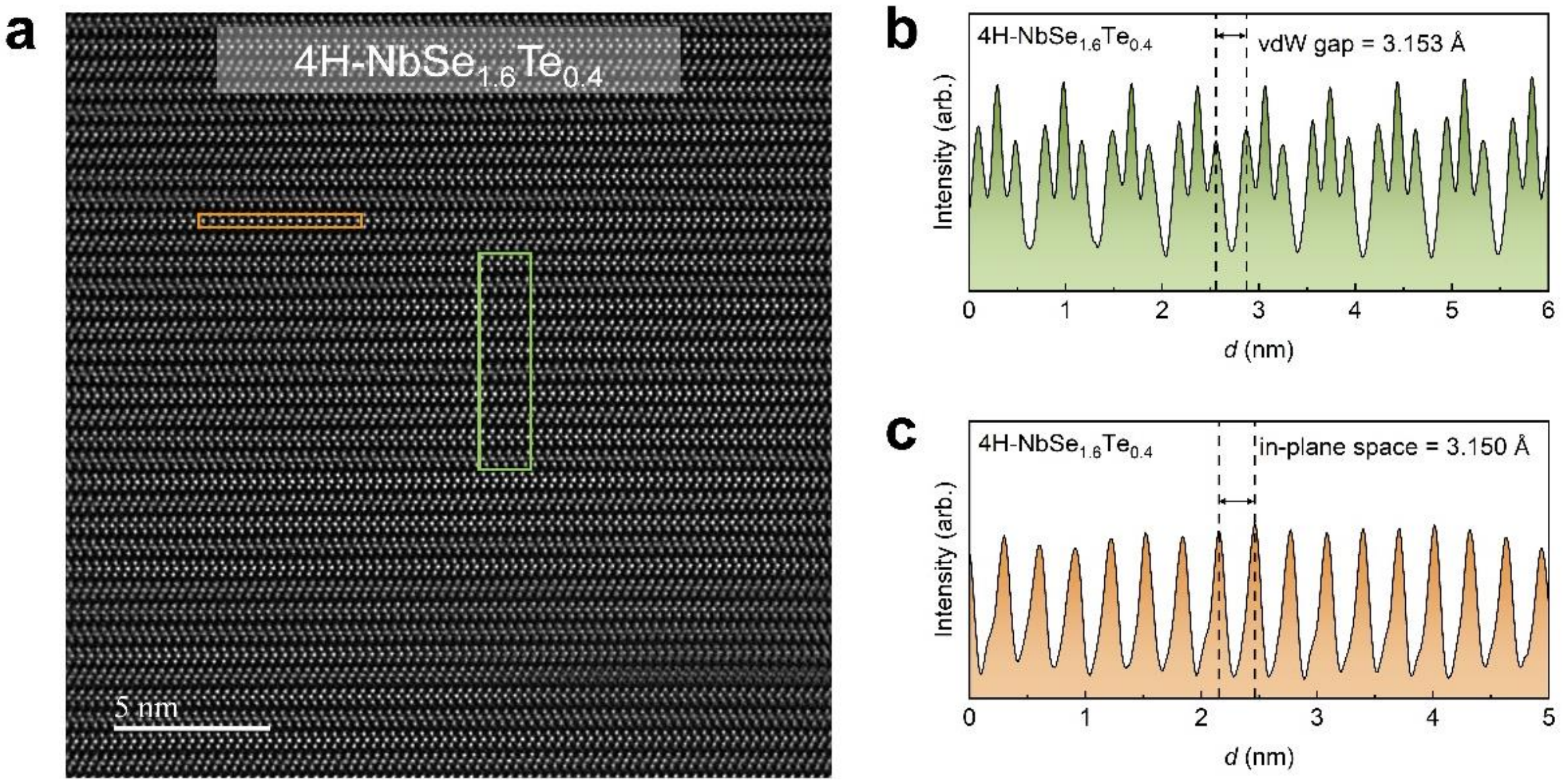

**Figure S7.** Analyses of vdW gap and projected in-plane Nb-column separation in 4H-$NbSe_{1.6}Te_{0.4}$. (a) Cross-sectional overview image of 4H-$NbSe_{1.6}Te_{0.4}$. (b,c) Corresponding signal intensity profiles, integrated from the selected regions, showing the atomic information along *c*-axis and *ab*-plane. Peak separations give the projected column distances along the selected profiles.

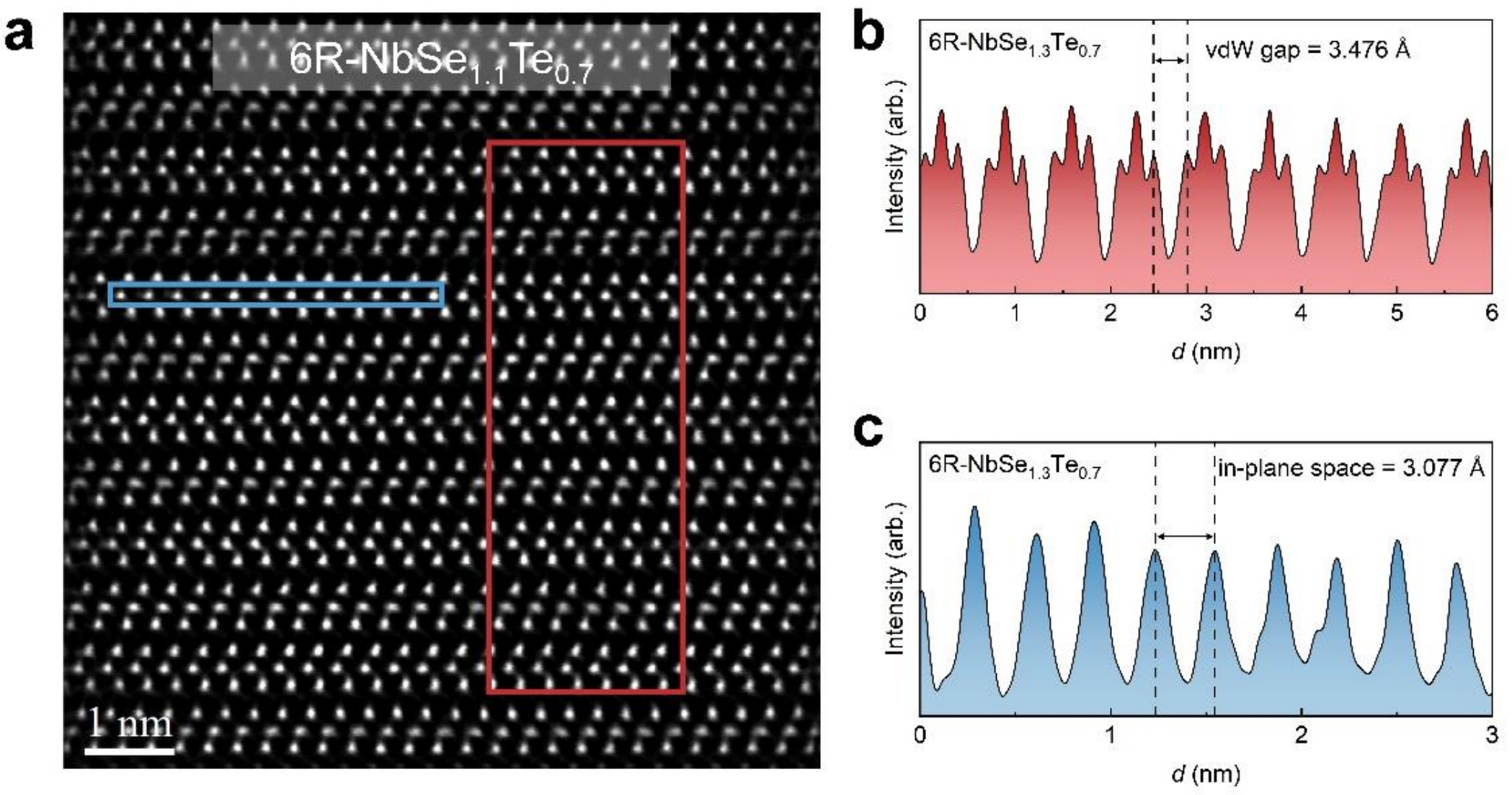

**Figure S8.** Analyses of vdW gap and projected in-plane Nb-column separation in 6R-$NbSe_{1.3}Te_{0.7}$. (a) Cross-sectional overview image of 6R-$NbSe_{1.3}Te_{0.7}$. (b,c) Corresponding signal intensity profiles, integrated from the selected regions, showing the atomic information along *c*-axis and *ab*-plane. Peak separations give the projected column distances along the selected profiles.

The vdW gaps measured in the displayed profiles are 3.153 Å for 4H, 3.476 Å for 6R and 2.925 Å for distorted 1T'. The corresponding in-plane profile annotations are 3.150, 3.077 and 3.153 Å (Figs. S7(c), S8(c) and S9(c)). These local projected distances describe the selected image regions, while Fig. 5(a) compares the diffraction-derived lattice parameters.

Projection geometry is important when comparing these quantities. For an ideal triangular Nb layer viewed along the three-index hexagonal [110] direction, adjacent projected column rows are separated by $\sqrt{3}a/2$. The local profiles and the diffraction lattice constants therefore probe different geometrical and spatial averages. We use the profile values to describe the local stacking geometry, and the diffraction results to discuss the average in-plane lattice evolution.

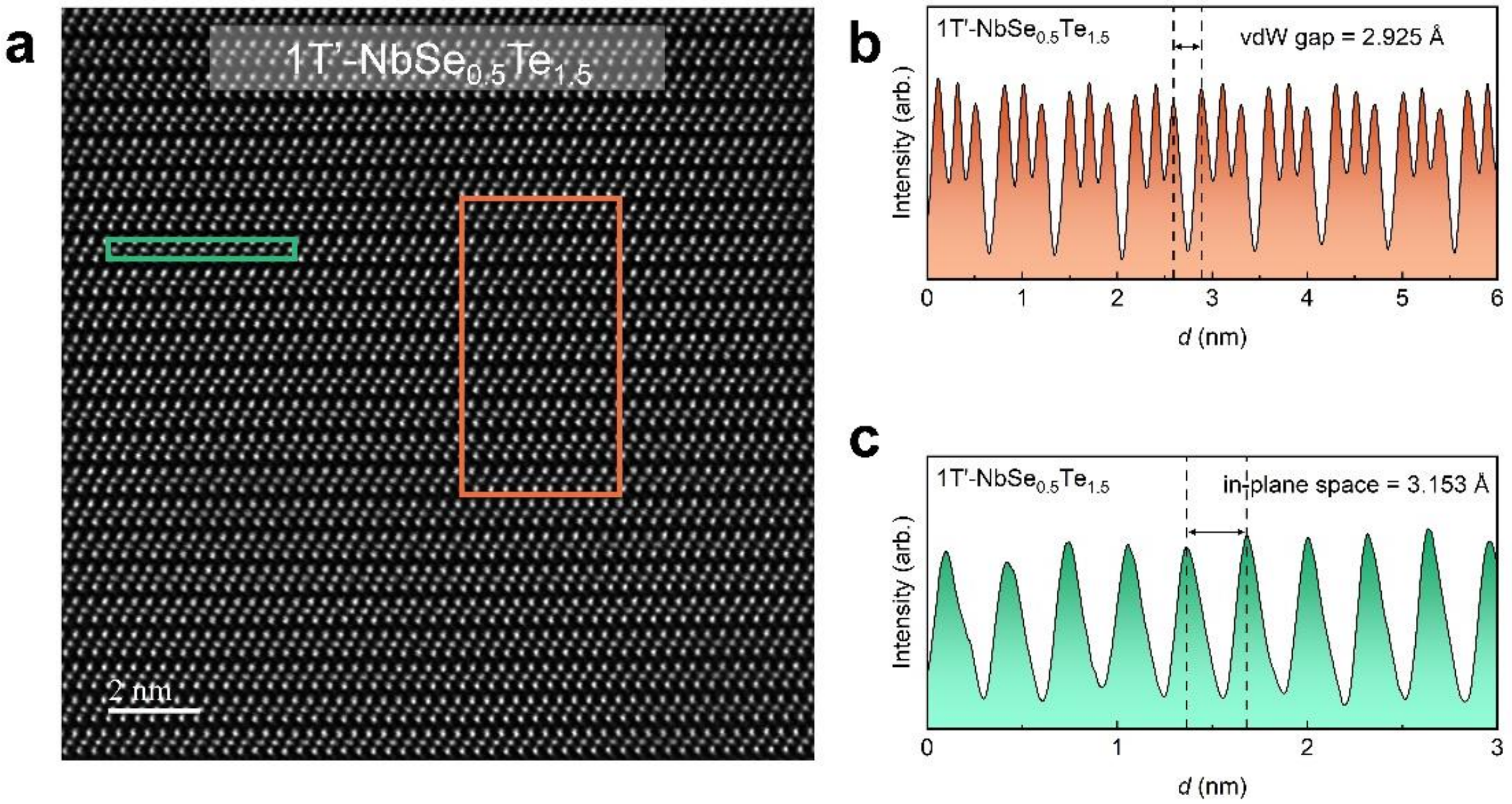

**Figure S9.** Analyses of vdW gap and projected in-plane Nb-column separation in 1T'-$NbSe_{0.5}Te_{1.5}$. (a) Cross-sectional overview image of 1T'-$NbSe_{0.5}Te_{1.5}$. (b,c) Corresponding signal intensity profiles, integrated from the selected regions, showing the atomic information along *c*-axis and *ab*-plane. Peak separations give the projected column distances along the selected profiles.

The contraction of the vdW gap in the distorted 1T' specimen accompanies the change to a T-derived stacking geometry and monoclinic coordination. The structural sequence therefore modifies both interlayer separation and local projected packing. Electronic coupling additionally depends on registry and orbital overlap, connecting these structural changes to the stacking-dependent electronic reconstruction.

## V. Characterizations of the electronic properties of $NbSe_{2-x}Te_x$ polytypes

For the transport data in Figs. S10-S16, the excitation current was 100 $\mu$A in the PPMS AC drive mode. The normal-state reference was $R_N = R_{xx}$(10 K). The zero-field $T_c$ and the upper critical fields were defined by the 50% $R_N$ criterion; the reduced-temperature plots use this $T_c$ for the corresponding specimen.

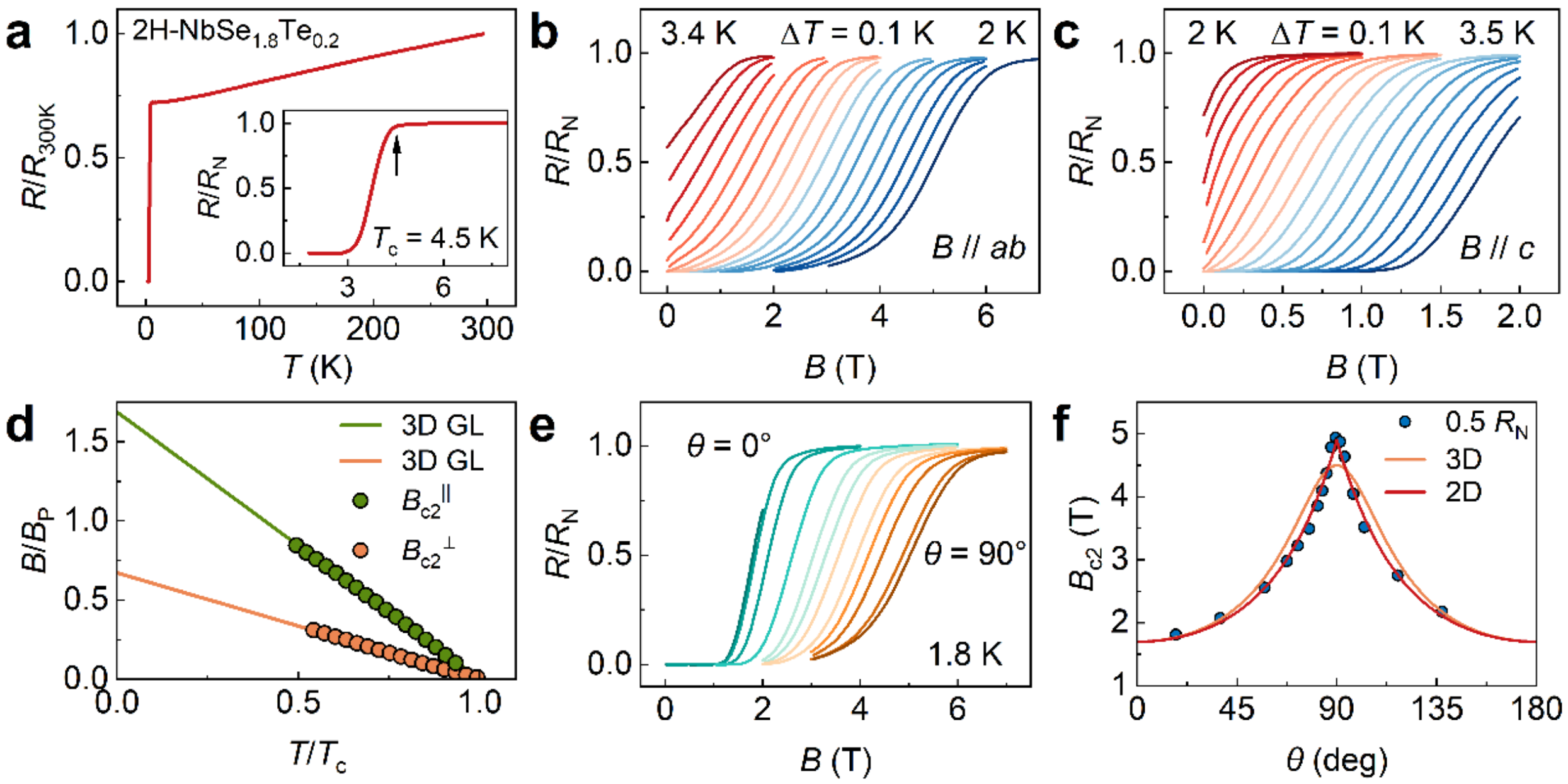

**Figure S10.** (a) $R(T)/R$(300 K) of 2H-$NbSe_{1.8}Te_{0.2}$; inset highlights $T_c$ = 4.5 K. (b, c) In-plane and out-of-plane field-dependent MR. (d) $B_{c2}(T)$ for $B//ab$-plane (green) and $B//c$ (yellow); solid lines: 3D GL fits[7,8]. (e) Angular MR from $ab$-plane ($\theta$ = 90°) to $c$-axis ($\theta$ = 0°). (f) $B_{c2}(\theta)$ at 1.8 K, fitted to the 2D Tinkham model[7,8].

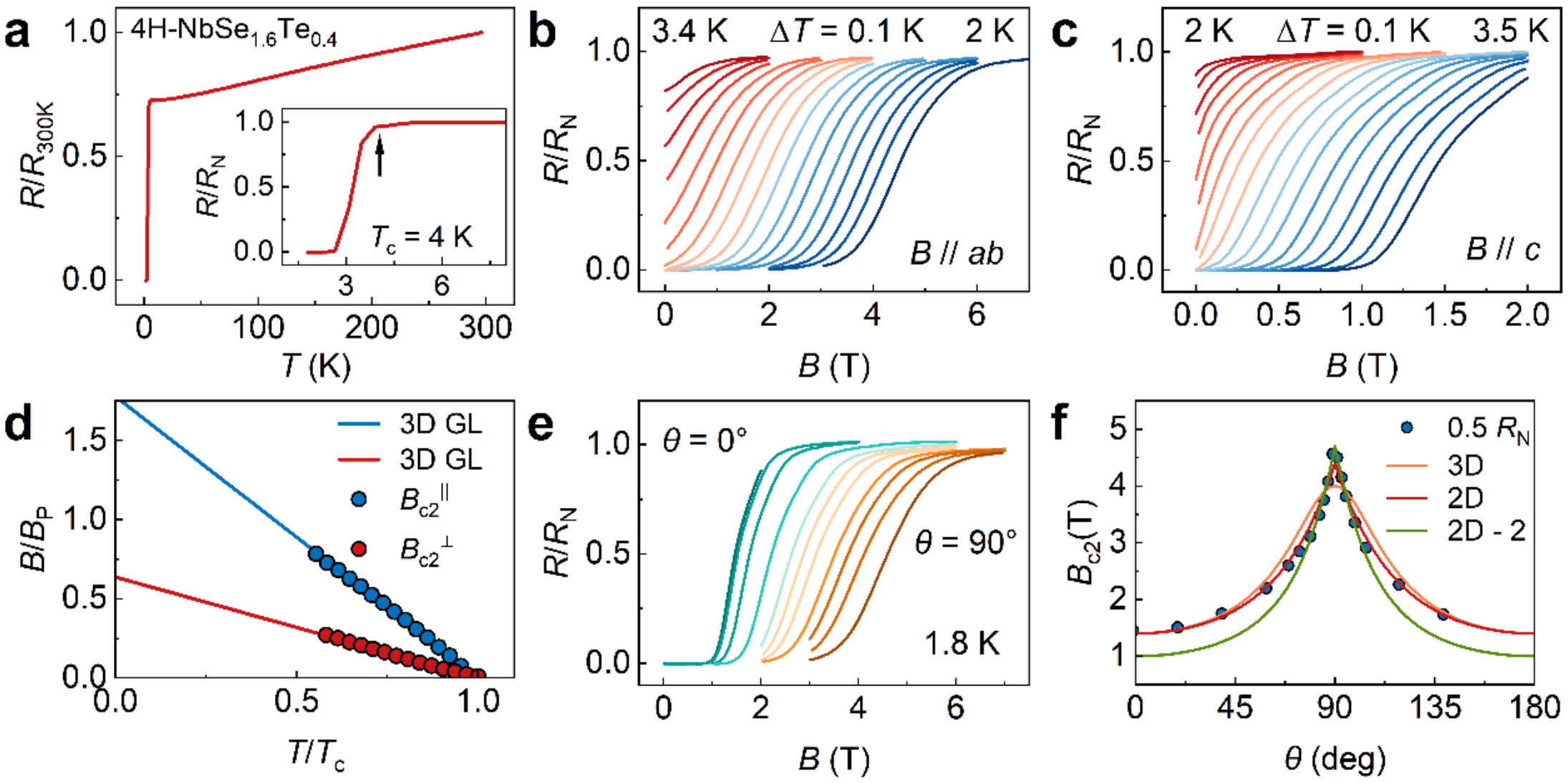

**Figure S11.** (a) $R(T)/R(300\text{ K})$ of 4H-$NbSe_{1.6}Te_{0.4}$; inset highlights $T_c = 4.0$ K. (b, c) In-plane and out-of-plane field-dependent MR. (d) $B_{c2}(T)$ for $B//ab$-plane (blue) and $B//c$ (red); solid lines: 3D GL fits[7,8]. (e) Angular MR from $ab$-plane ($\theta = 90°$) to $c$-axis ($\theta = 0°$). (f) $B_{c2}(\theta)$ at 1.8 K, showing a notable enhancement for in-plane fields, fitted by two distinct 2D Tinkham models[7-13].

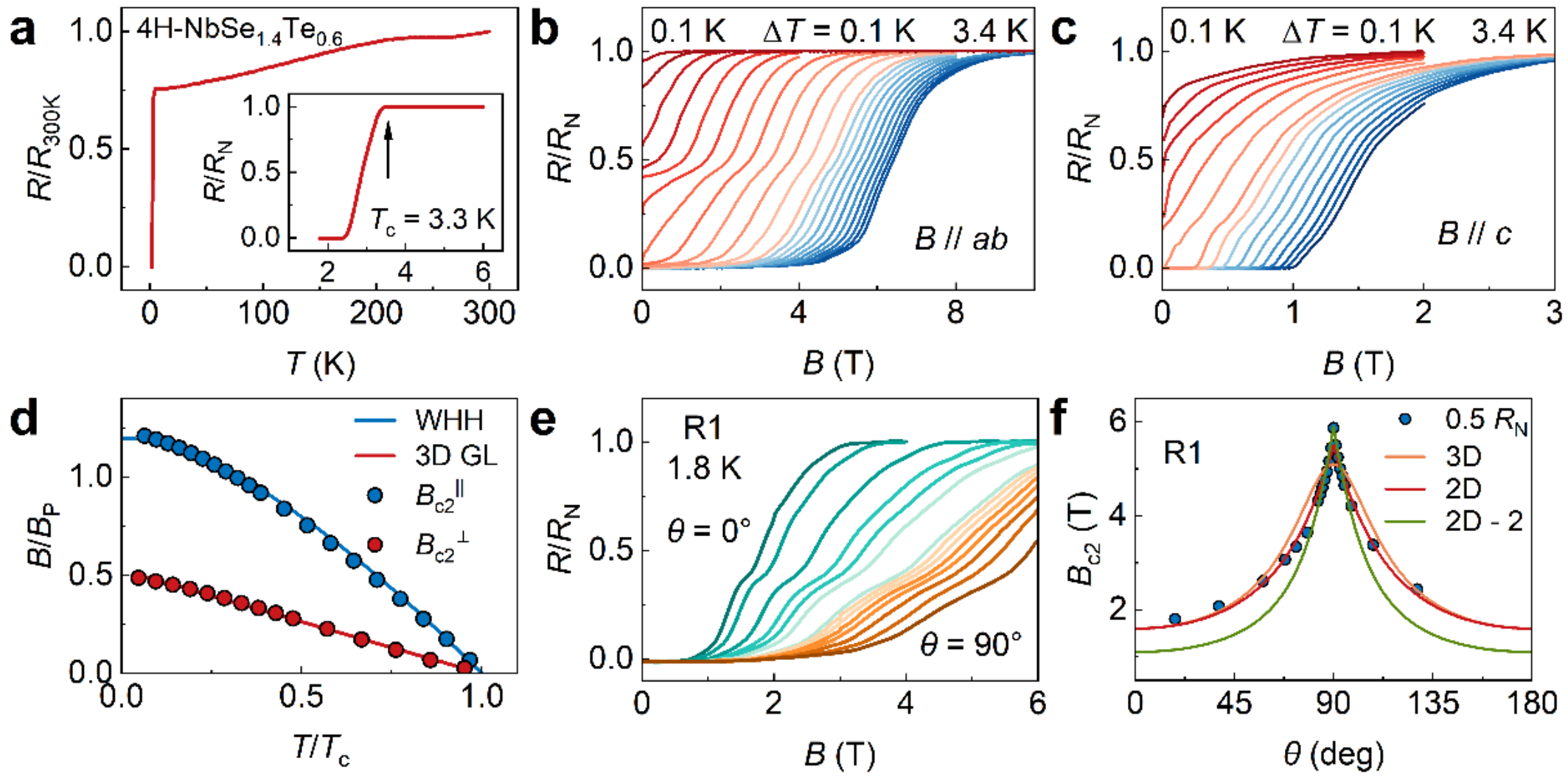

**Figure S12.** (a) $R(T)/R(300\ \mathrm{K})$ of 4H-$NbSe_{1.4}Te_{0.6}$; inset: superconducting transition at 3.3 K. (b,c) Field-dependent resistance for $B//ab$ and $B//c$. (d) $B_{c2}(T)$ normalized by the Pauli reference, with WHH ($\alpha = 5.2$, $\lambda_{so} = 7.1$)[14] and GL descriptions for the parallel and perpendicular fields, respectively. (e) Angular resistance and (f) $B_{c2}(\theta)$, showing the broad two-dimensional cusp and the additional near-plane enhancement described by two local Tinkham curves[7,8].

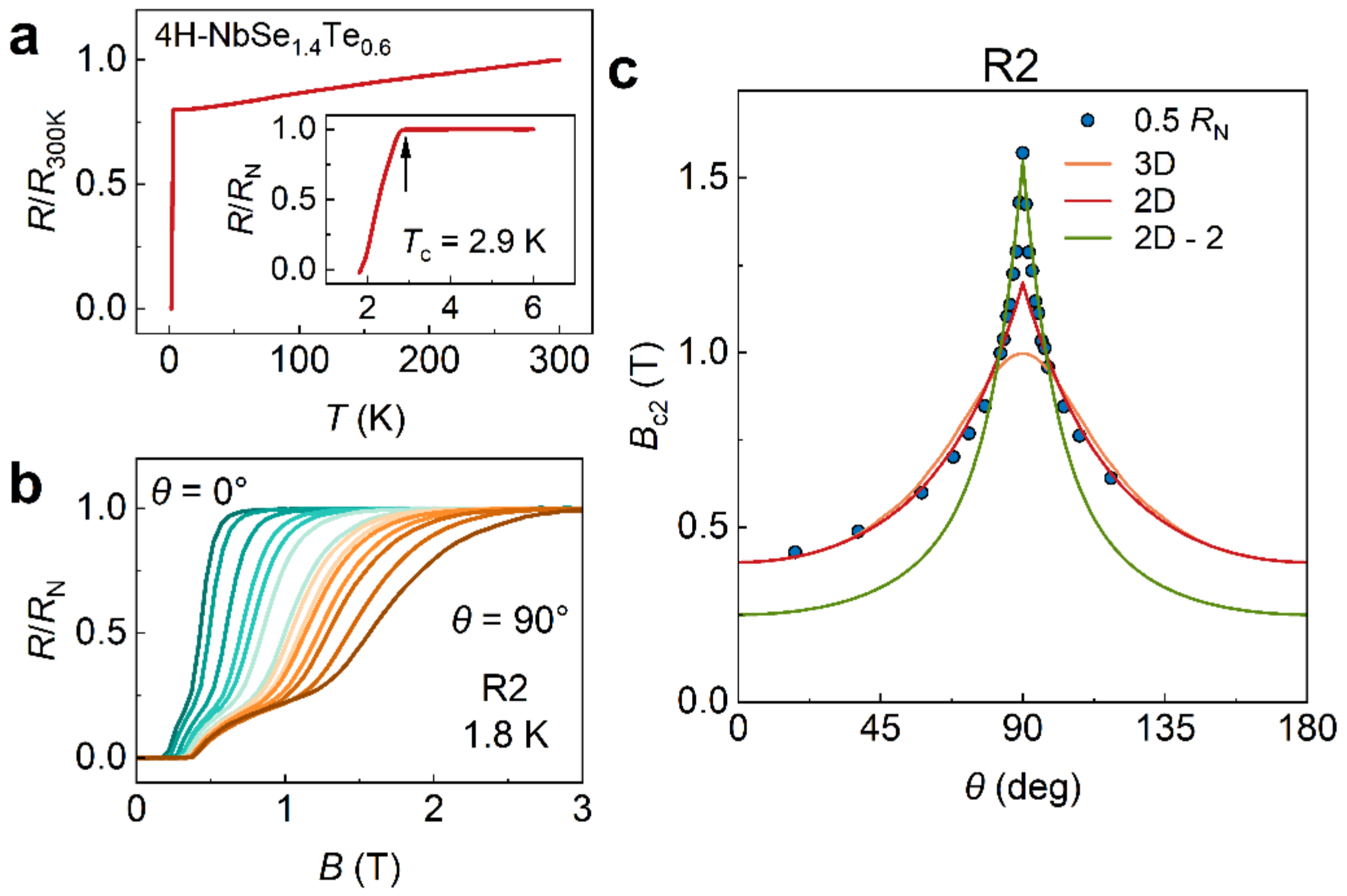

**Figure S13.** (a) $R(T)/R(300\ \mathrm{K})$ of a second 4H-$NbSe_{1.4}Te_{0.6}$ crystal (R2) from an independent batch; inset highlights $T_c \approx 3$ K. (b) Angular MR from *ab*-plane ($\theta = 90°$) to *c*-axis ($\theta = 0°$). (c) $B_{c2}(\theta)$ at 1.8 K, reproducing the in-plane enhancement and fitted by two local 2D Tinkham descriptions[7-13], confirming the repeatability of the anisotropic behavior.

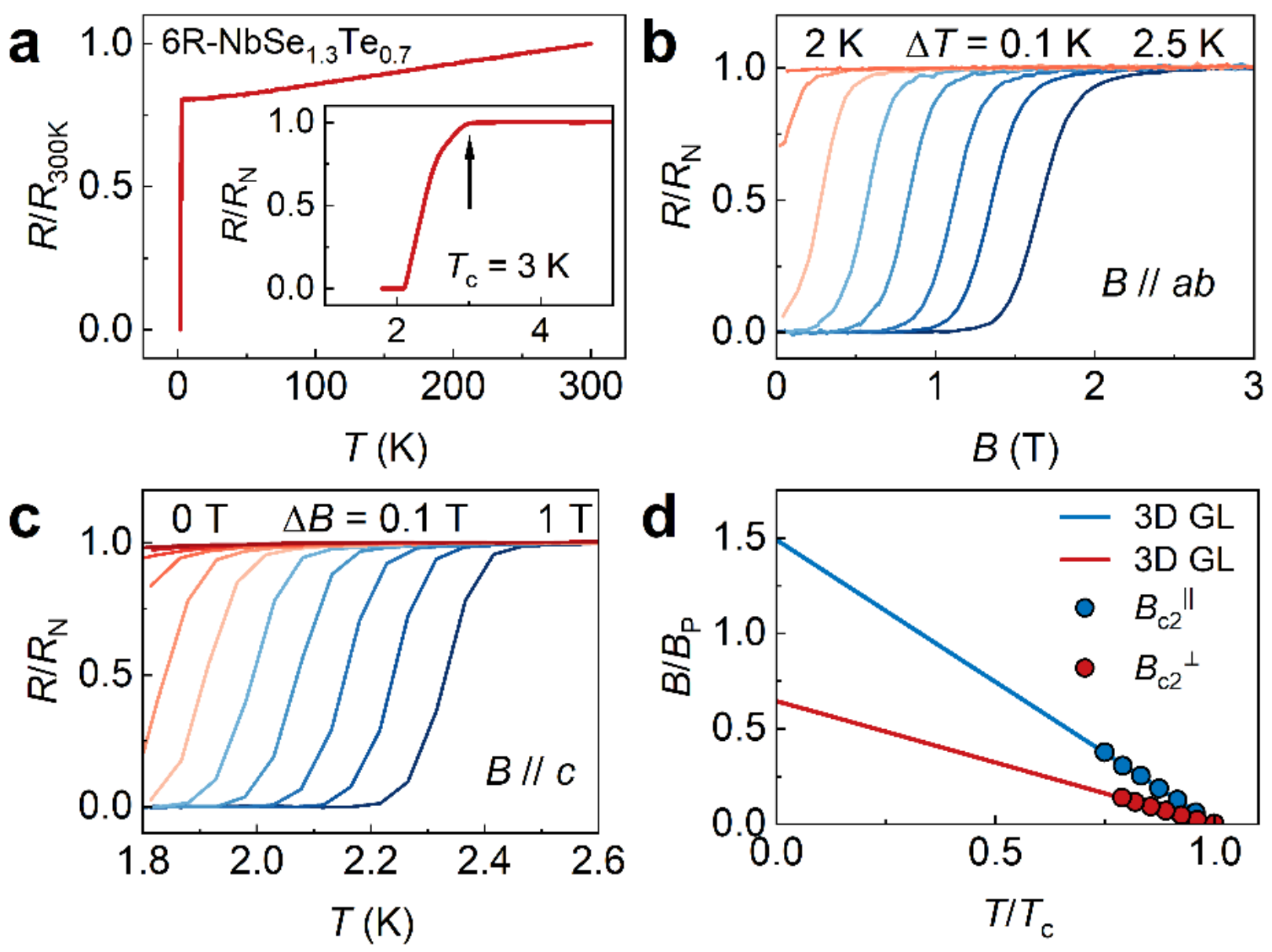

**Figure S14.** (a) $R(T)/R(300\text{ K})$ of 6R-$NbSe_{1.3}Te_{0.7}$; inset highlights $T_c = 3.0$ K. (b, c) In-plane and out-of-plane field-dependent MR. (d) $B_{c2}(T)$ for $B//ab$-plane (blue) and $B//c$ (red); solid lines: 3D GL fits[7,8].

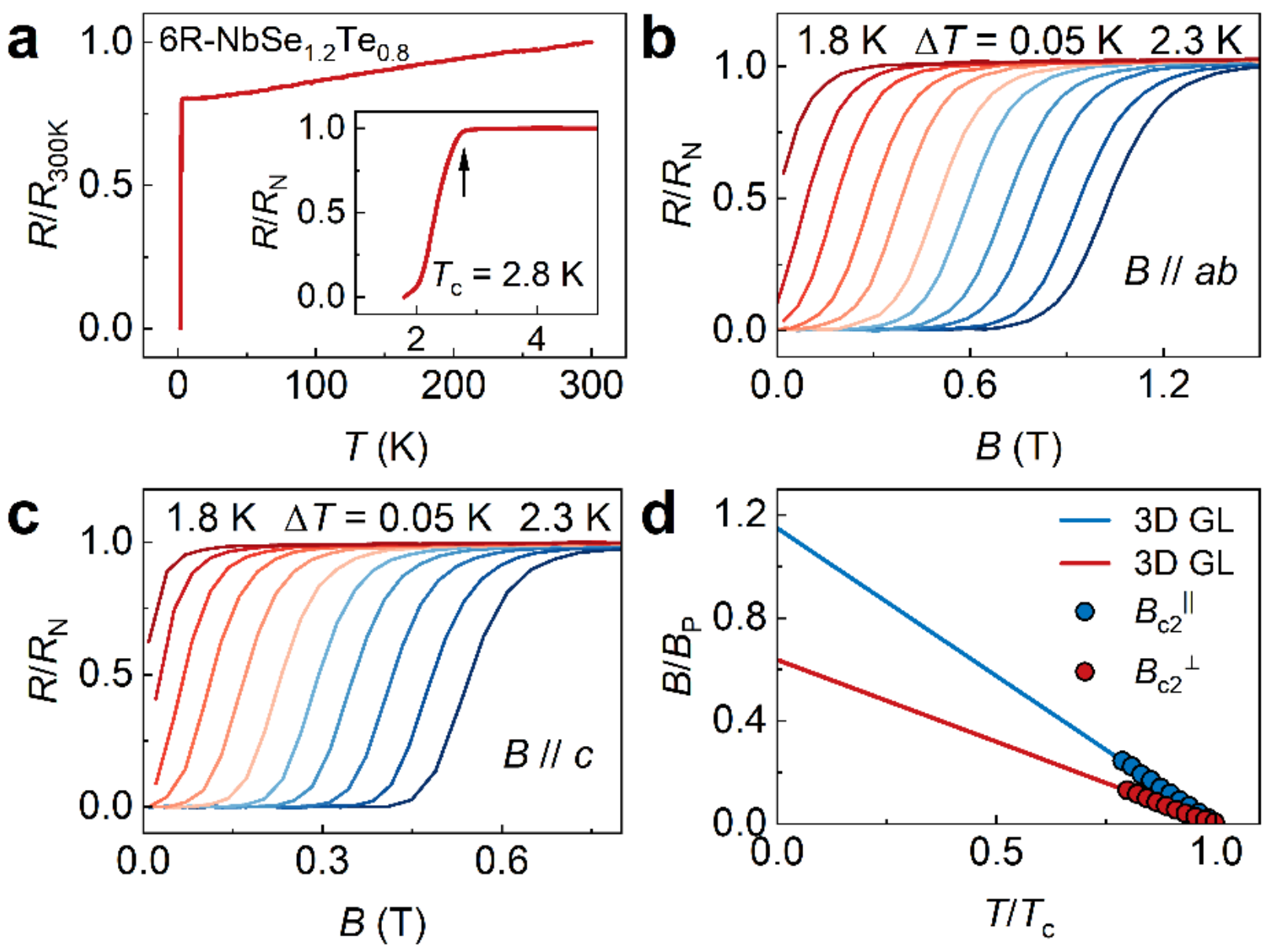

**Figure S15.** (a) $R(T)/R(300\ \mathrm{K})$ of the 6R-$NbSe_{1.2}Te_{0.8}$ specimen in Fig. 2; inset highlights $T_c$ = 2.8 K. (b, c) In-plane and out-of-plane field-dependent MR. (d) $B_{c2}(T)$ for *B*//*ab*-plane (blue) and *B*//*c* (red); solid lines: 3D GL fits[7,8].

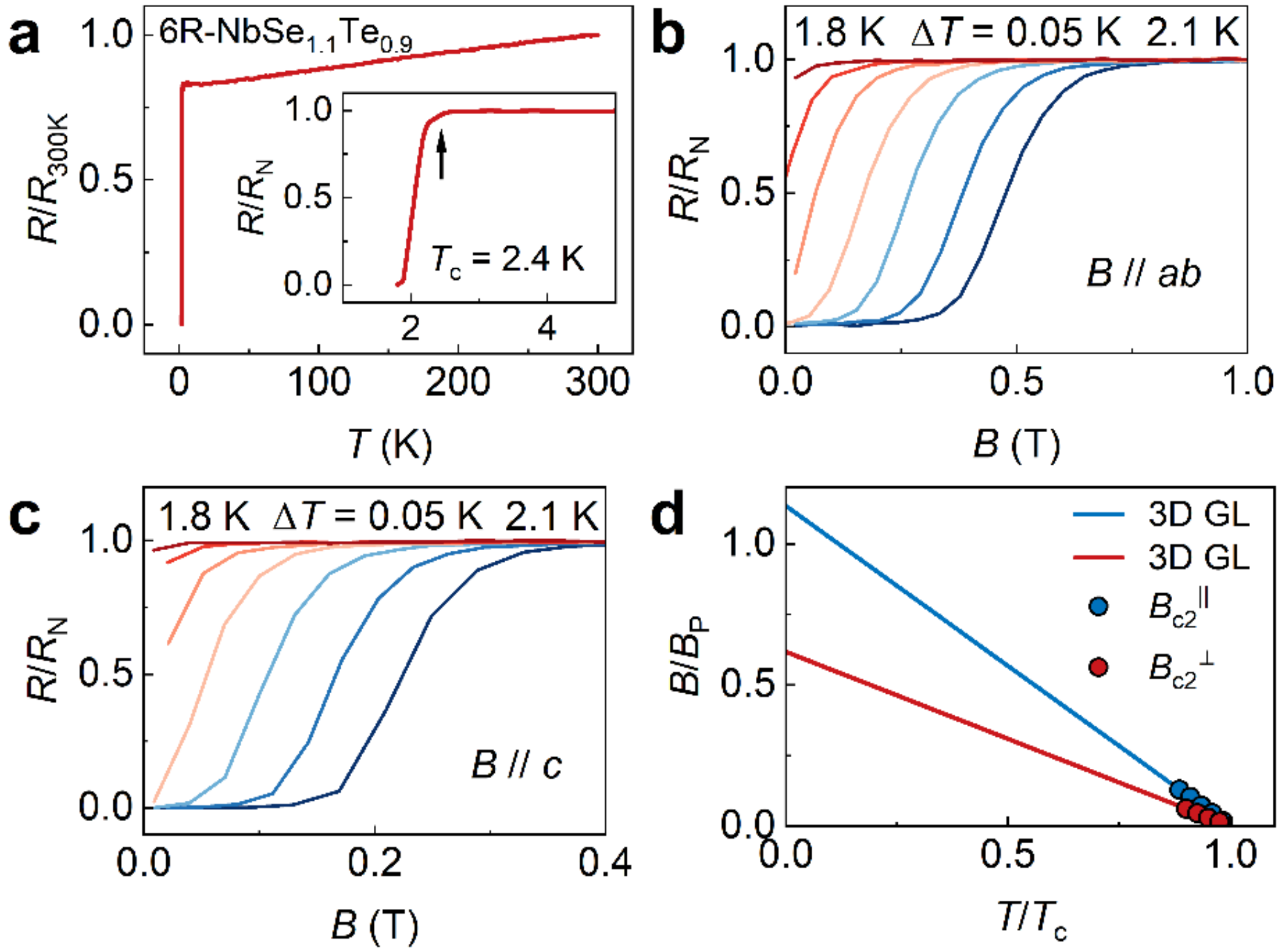

**Figure S16.** (a) $R(T)/R(300\ \mathrm{K})$ of 6R-$NbSe_{1.1}Te_{0.9}$; inset highlights $T_c$ = 2.4 K. (b, c) In-plane and out-of-plane field-dependent MR. (d) $B_{c2}(T)$ for *B*//*ab*-plane (blue) and *B*//*c* (red); solid lines: 3D GL fits[7,8].

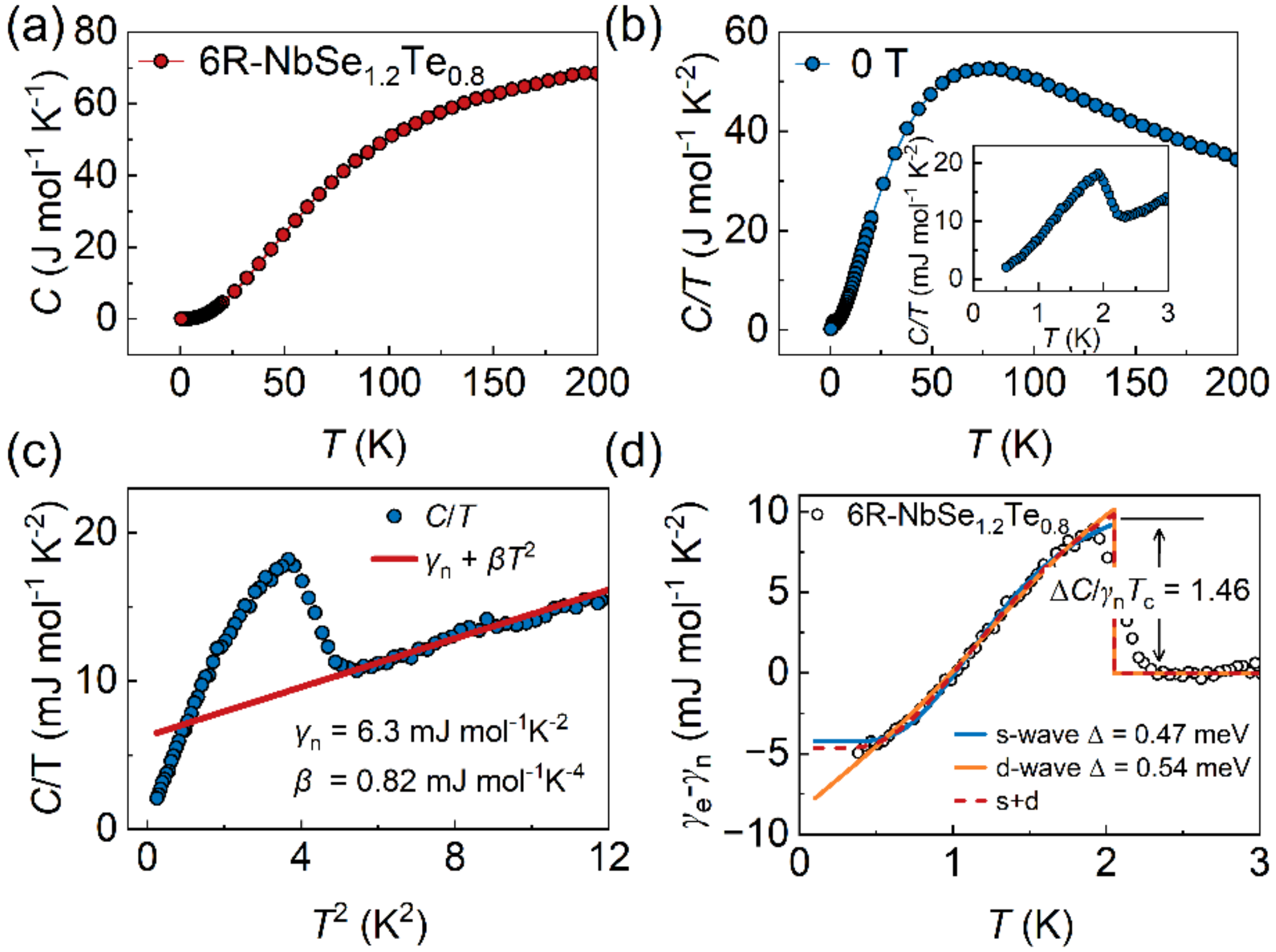

**Figure S17.** (a) Zero-field specific heat $C(T)$ of 6R-$NbSe_{1.2}Te_{0.8}$. (b) $C/T$ versus $T$; inset: superconducting anomaly. (c) Normal-state fit $C/T = \gamma_n + \beta T^2$, giving $\gamma_n = 6.3$ mJ mol$^{-1}$ K$^{-2}$ and $\beta$ = 0.82 mJ mol$^{-1}$ K$^{-4}$. (d) Electronic specific-heat coefficient relative to its normal-state value, $C_e/T - \gamma_n$, compared with *s*-wave (blue), *d*-wave (orange) and two-component *s*+*d*-wave (red dashed) models. The *s*+*d* response is $C_e(T) = wC_{e,s}(T) + (1 - w)C_{e,d}(T)$, with the component functions and their weight defined in Methods. The jump $\Delta C/(\gamma_n T_c) \approx 1.46$ is close to the weak-coupling BCS value of 1.43[15,16].

The specific heat of 6R-$NbSe_{1.2}Te_{0.8}$ provides thermodynamic evidence for bulk superconductivity (Fig. S17). The zero-field data show an anomaly near 2.1 K. Subtraction of the phonon term obtained from $C/T = \gamma_n + \beta T^2$ gives the electronic contribution, with $\Delta C/(\gamma_n T_c) \approx 1.46$.

Figure S17(d) compares the *s*-, *d*- and two-component *s*+*d*-wave models defined in Methods. The weighted *s*+*d* response follows the temperature dependence below the transition more closely than the isotropic *s*-wave comparison. The *d*-wave component supplies an angle-dependent contribution within this thermodynamic model. Related specific-heat studies illustrate how multiple gap contributions modify the electronic heat capacity[17-19].

## VI. Theoretical calculations

The relative-energy calculations are designed to capture the composition-driven competition among different stacking motifs, rather than to identify a unique ground-state structure for each nominal composition. At a fixed value of $x$, $NbSe_{2-x}Te_x$ remains a substitutional alloy: the same average Te concentration can be realized through numerous inequivalent local Se/Te arrangements within the finite DFT supercell[20-27]. This configurational freedom gives rise to the multiple energy points shown for each structural type in Fig. S18. Consequently, an isolated low-energy point for a competing phase should be interpreted as a possible local substitution pattern, not as definitive evidence that this phase must be obtained experimentally. The finite set of configurations does not establish convergence of the minimum energy with configuration number or demonstrate an exhaustive ground-state search; the calculated ordering therefore does not provide a quantitative prediction of the experimental phase boundaries.

Near $x = 0.83$, the sampled 6R configurations compete closely with distorted 1T' configurations. A possible star-of-David reconstruction of the T layers provides an additional energy-lowering mechanism for the H/T superlattice, as calculated for a simplified model in Ref. 2. The formation energies in Figs. 5(e) and S18 describe the unreconstructed configurations. The measured in-plane resistance curves in Figs. S14–S16 show no resolved CDW-like kink or upturn. Random Se/Te potentials and local stacking deviations can disrupt the spatial coherence of long-range charge order. High-temperature CVT growth followed by rapid quenching may retain locally disordered or metastable stacking arrangements, broadening or weakening the CDW transition signature. Short-range charge correlations remain compatible with this smooth transport. The finite configurational sampling and local Se/Te environments also influence the relative energies.

Near $x$ = 1.0, the 6R, 1T and distorted 1T' configurations are close in energy. Local Se/Te arrangements, electronic correlations and reconstruction of the T-derived layers can influence this competition. High-temperature CVT growth followed by quenching also permits metastable stacking sequences to be retained. The near-degenerate energy landscape and the configurational freedom of the alloy explain why the experimentally obtained sequence can differ from the single lowest-energy finite-supercell configuration.

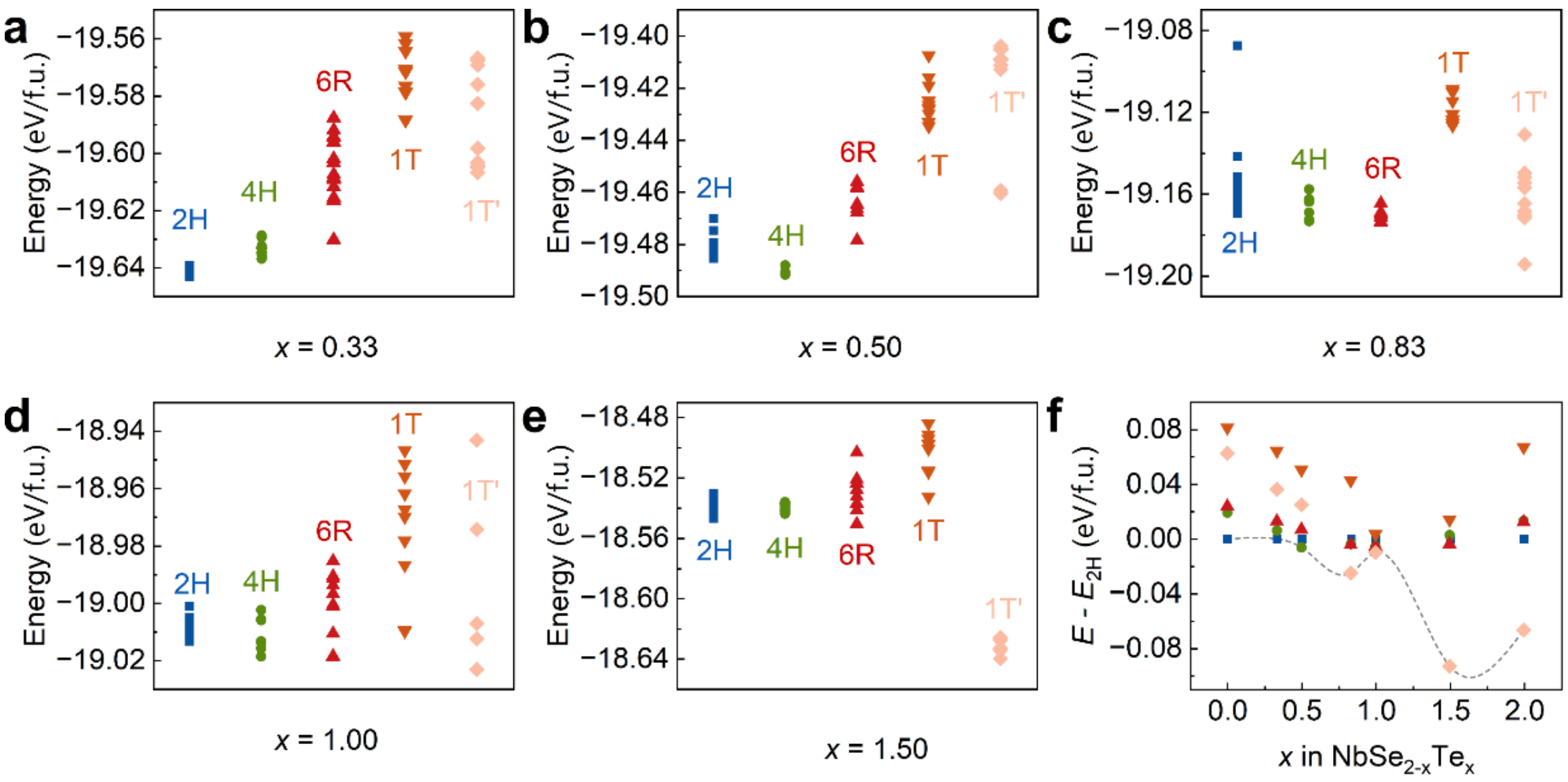

**Figure S18.** (a-e) Formation energies of 2H, 4H, 6R, 1T, and distorted 1T' phases for $x$ = 0.33 to 1.50; multiple points reflect inequivalent Se/Te substitution sites. (f) Relative formation energies referenced to the 2H phase at each $x$. The dashed line connects the lowest-energy configurations, guiding the phase competition analysis.

**SOC resolved electronic structures**

To examine the role of SOC in the electronic structure, we calculated 4H-$NbSe_{1.4}Te_{0.6}$ and 6R-$NbSe_{1.3}Te_{0.7}$ using the virtual crystal approximation. Figures S19 and S20 compare the dispersions with and without SOC in the $k_z = 0$ and $k_z = \pi/c$ planes, respectively. SOC modifies the bands most visibly along K-Γ/H-A and M-K/L-H. The out-of-plane spin projections are shown in Figs. S21 and S22.

In the $R3m$ 6R structure, SOC produces opposite-$S_z$ branches with a characteristic splitting of approximately 0.185 eV near K/H. The $P6_3/mmc$ 4H structure preserves inversion symmetry and the bulk bands remain doubly degenerate. This distinction concerns the global bands: inversion-asymmetric H layers can retain local SOC with opposite layer textures in a centrosymmetric stacking, as established for 2H-$NbSe_2$. SOC and interlayer hybridization therefore enter the superconducting response together; a larger global spin splitting alone does not imply stronger superconducting two-dimensionality.[28]

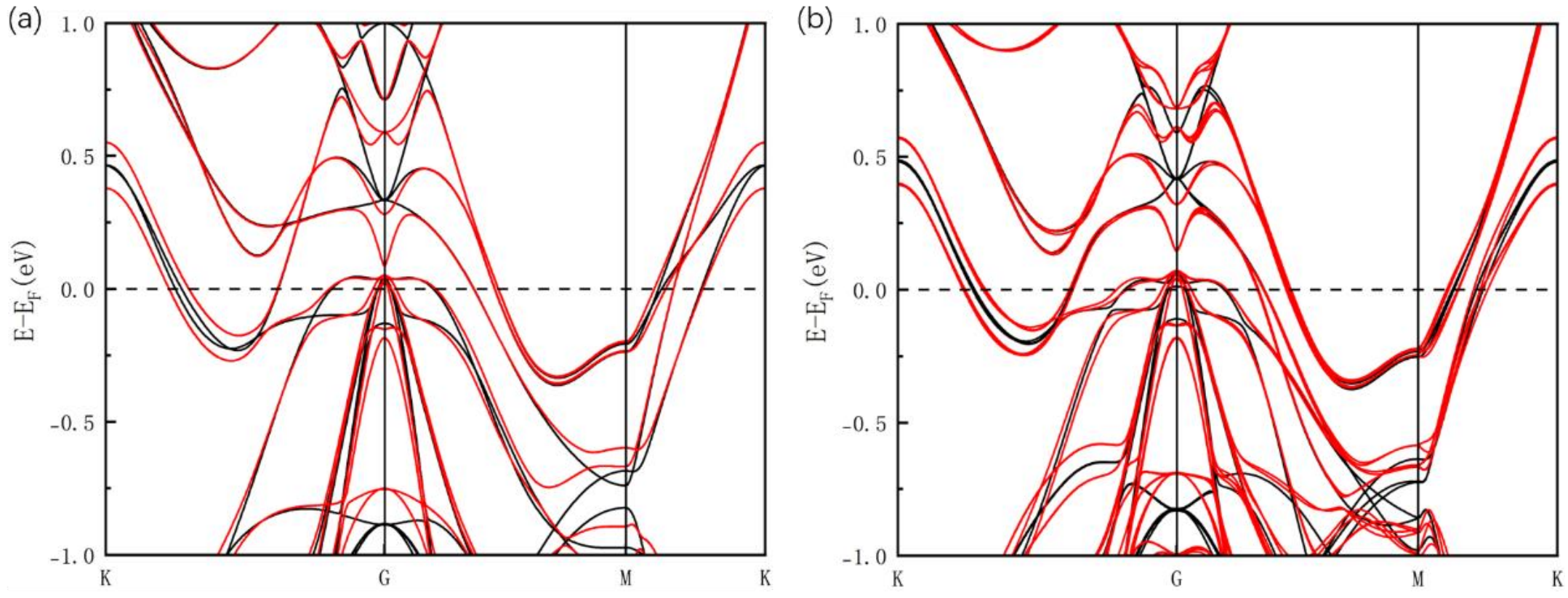

**Figure S19.** Calculated band structures of (a) 4H-$NbSe_{1.4}Te_{0.6}$ and (b) 6R-$NbSe_{1.3}Te_{0.7}$ along K-Γ-M-K in the $k_z = 0$ plane. Red and black curves denote calculations with and without spin–orbit coupling, respectively. The Se/Te substitution is treated using the virtual crystal approximation. The Fermi energy is set to zero.

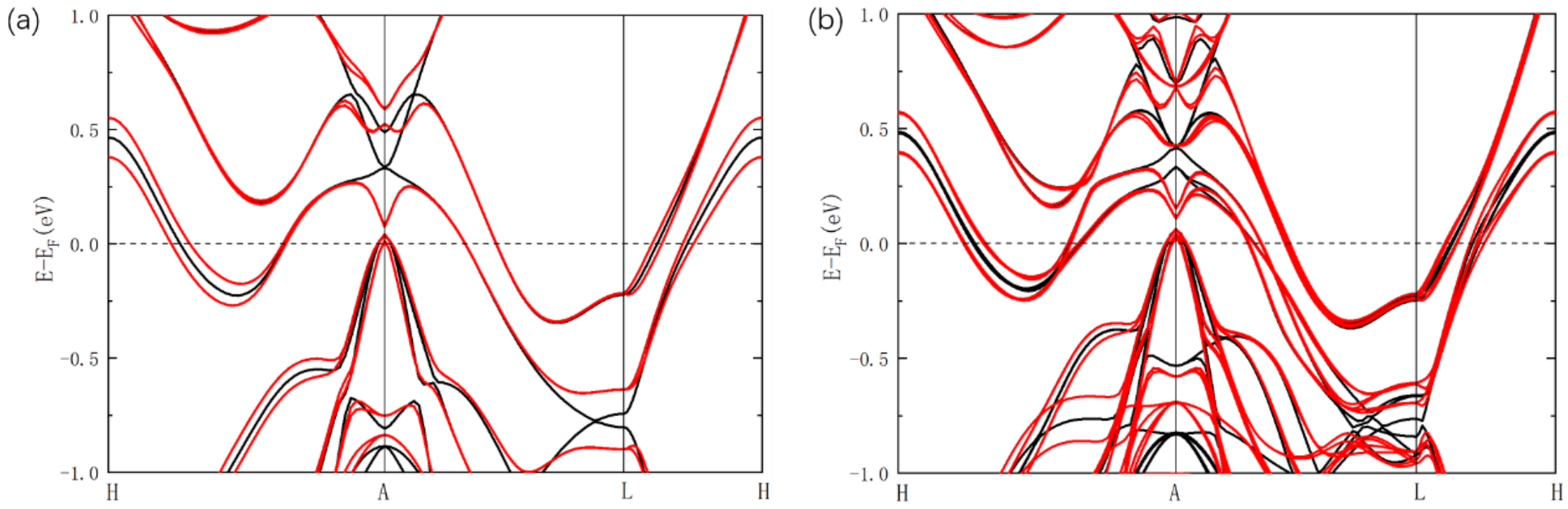

**Figure S20.** Calculated band structures of (a) 4H-$NbSe_{1.4}Te_{0.6}$ and (b) 6R-$NbSe_{1.3}Te_{0.7}$ along H-A-L-H in the $k_z = \pi/c$ plane. Red and black curves denote calculations with and without spin–orbit coupling, respectively. The Se/Te substitution is treated using the virtual crystal approximation. The Fermi energy is set to zero.

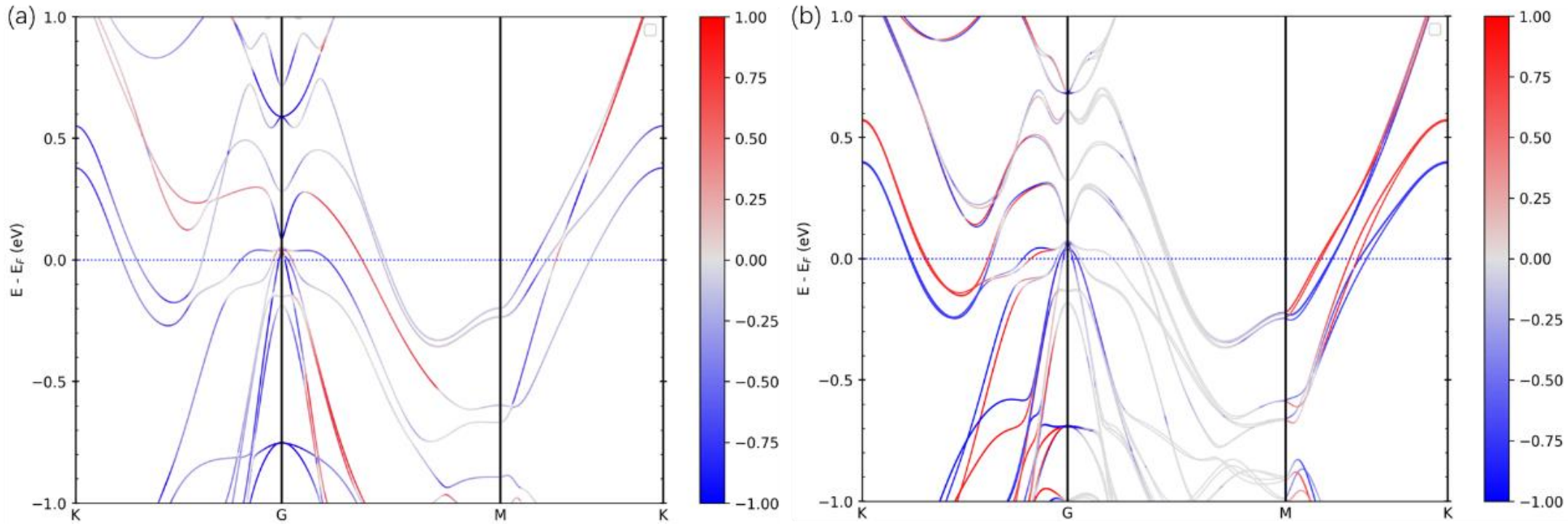

**Figure S21.** Spin–orbit-coupled band structures of (a) 4H-$NbSe_{1.4}Te_{0.6}$ and (b) 6R-$NbSe_{1.3}Te_{0.7}$ along K-Γ-M-K at $k_z = 0$, colored by the out-of-plane spin projection $S_z$. The Fermi energy is set to zero.

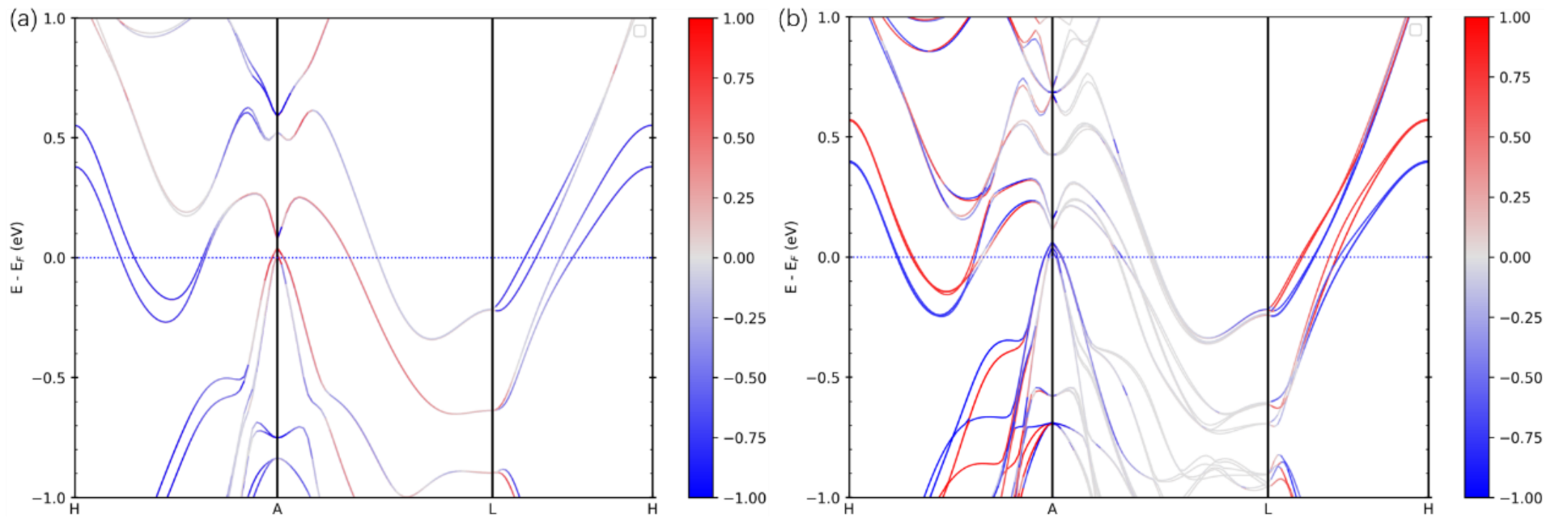

**Figure S22.** Spin–orbit-coupled band structures of (a) 4H-$NbSe_{1.4}Te_{0.6}$ and (b) 6R-$NbSe_{1.3}Te_{0.7}$ along H-A-L-H at $k_z = \pi/c$, colored by the out-of-plane spin projection $S_z$. The Fermi energy is set to zero.

## VII Directional normal state transport

Normal-state transport provides a complementary comparison of the layer anisotropy (Fig. S23). For representative 2H-$NbSe_{1.8}Te_{0.2}$, 4H-$NbSe_{1.6}Te_{0.4}$ and 6R-$NbSe_{1.1}Te_{0.9}$ crystals, $\rho_{zz}/\rho_{xx}$ is enhanced toward low temperature and reaches its largest values in 4H. The increase from 2H to 4H and decrease toward 6R agree with the evolution of superconducting two-dimensionality. This correspondence supports the role of stacking-dependent interlayer charge motion in the superconducting anisotropy.

The $c$-axis resistivity was measured in a Corbino geometry with current along $c$. Figure S23 compares $\rho_{zz}/\rho_{xx}$ for representative crystals of the three H-containing polytypes[29]. The largest resistivity anisotropy occurs in 4H, consistent with its enhanced superconducting anisotropy.

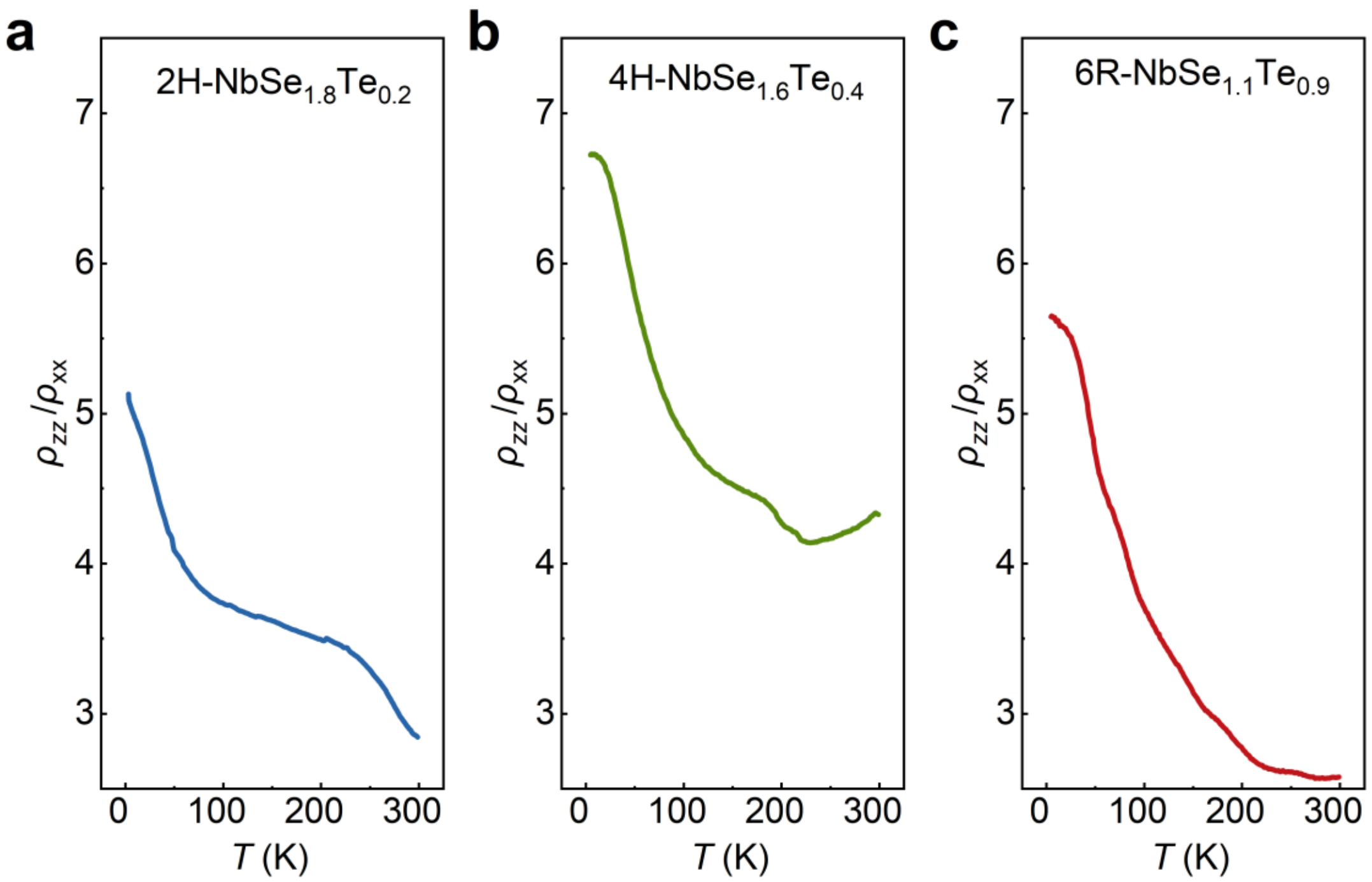

**Figure S23.** Temperature dependence of the normal-state resistivity ratio $\rho_{zz}/\rho_{xx}$ for representative (a) 2H-$NbSe_{1.8}Te_{0.2}$, (b) 4H-$NbSe_{1.6}Te_{0.4}$ and (c) 6R-$NbSe_{1.1}Te_{0.9}$ crystals. The *c*-axis component $\rho_{zz}$ was measured in a Corbino geometry; $\rho_{xx}$ denotes the in-plane resistivity.

**Table S1. Crystallographic data of 4H-$NbSe_{1.6}Te_{0.4}$ at 293 K using the centrosymmetric $P6_3/mmc$ structure. Nb occupancies were fixed at 1.0 during refinement.**

| | |
|---|---|
| Chemical formula | $NbSe_{1.6}Te_{0.4}$ |
| Formula weight | 270.29 g/mol |
| Temperature | 293.00 K |
| Crystal system | Hexagonal |
| Space group | $P6_3/mmc$ |
| | $a$ = 3.5213(4) Å, $\alpha$ = 90° |
| Unit-cell dimensions | $b$ = 3.5213(4) Å, $\beta$ = 90° |
| | $c$ = 25.9219(24) Å, $\gamma$ = 120° |

| Label | $x$ | $y$ | $z$ | Occupancy | Wyckoff |
|---|---|---|---|---|---|
| Nb 1 | 1 | 0 | 0.25000 | 1.0 | 2*b* |
| Nb 2 | 0 | 1 | 0.50000 | 1.0 | 2*a* |
| Se/Te 1 | 1/3 | 2/3 | 0.43039 | 0.8/0.2 | 4*f* |
| Se/Te 2 | 2/3 | 1/3 | 0.31820 | 0.8/0.2 | 4*f* |

**Table S2. Crystallographic data of 6R-$NbSe_{1.1}Te_{0.9}$ at 293 K by XRD refinement. Nb occupancies were fixed at 1.0 during refinement.**

| | |
|---|---|
| Chemical formula | $NbSe_{1.1}Te_{0.9}$ |
| Formula weight | 294.60 g/mol |
| Temperature | 293.00 K |
| Crystal system | Rhombohedral |
| Space group | $R3m$ |
| Unit-cell dimensions | $a$ = 3.5240(5) Å, $\alpha$ = 90°<br>$b$ = 3.5240(5) Å, $\beta$ = 90°<br>$c$ = 39.2434(27) Å, $\gamma$ = 120° |

| Label | $x$ | $y$ | $z$ | Occupancy | Wyckoff |
|---|---|---|---|---|---|
| Nb 1 | 0 | 2/3 | 0.58325 | 1.0 | 3*a* |
| Nb 2 | 2/3 | 1/3 | 0.41665 | 1.0 | 3*a* |
| Se/Te 1 | 0 | 0 | 0.53772 | 0.55/0.45 | 3*a* |
| Se/Te 2 | 0 | 0 | 0.62744 | 0.55/0.45 | 3*a* |
| Se/Te 3 | 0 | 0 | 0.37049 | 0.55/0.45 | 3*a* |
| Se/Te 4 | 1/3 | 2/3 | 0.46136 | 0.55/0.45 | 3*a* |

**Table S3. Crystallographic data of 1T'-$NbSe_{0.5}Te_{1.5}$ at 293 K by XRD refinement. Nb occupancies were fixed at 1.0 during refinement.**

| | |
|---|---|
| Chemical formula | $NbSe_{0.5}Te_{1.5}$ |
| Formula weight | 323.79 g/mol |
| Temperature | 293.00 K |
| Crystal system | Monoclinic |
| Space group | *C*2/*m* |
| Unit-cell dimensions | $a$ = 14.593(10) Å, $\alpha$ = 90°<br>$b$ = 10.7966(18) Å, $\beta$ = 110.555(13)°<br>$c$ = 9.210(4) Å, $\gamma$ = 90° |

| Label | *x* | *y* | *z* | Occupancy | Wyckoff |
|---|---|---|---|---|---|
| Nb 1 | 0.50000 | 0.00000 | 0.50000 | 1.0 | 2*d* |
| Nb 2 | 0.34713 | 0.16218 | 0.19573 | 1.0 | 8*j* |
| Nb 3 | 0.50000 | 0.33511 | 0.50000 | 1.0 | 4*h* |
| Nb 4 | 0.35335 | 0.50000 | 0.17238 | 1.0 | 4*i* |
| Se/Te 1 | 0.35463 | 0.16602 | 0.49152 | 0.25/0.75 | 8*j* |
| Se/Te 2 | 0.48459 | 0.00000 | 0.19260 | 0.25/0.75 | 4*i* |
| Se/Te 3 | 0.34925 | 0.50000 | 0.49837 | 0.25/0.75 | 4*i* |
| Se/Te 4 | 0.18738 | 0.32626 | 0.12152 | 0.25/0.75 | 8*j* |
| Se/Te 5 | 0.49667 | 0.33724 | 0.18420 | 0.25/0.75 | 8*j* |
| Se/Te 6 | 0.19571 | 0.00000 | 0.15610 | 0.25/0.75 | 4*i* |